\documentclass[11pt]{article}

\usepackage{graphicx}
  
\usepackage{multicol} 

\usepackage{color}
\usepackage{latexsym}

\definecolor{rosso}{cmyk}{0,1,1,0.4}
\definecolor{rossos}{cmyk}{0,1,1,0.55}
\definecolor{rossoc}{cmyk}{0,0.5,1,0.2}
\definecolor{blu}{cmyk}{1,1,0,0.3}
\definecolor{blus}{cmyk}{1,1,0,0.6}
\definecolor{blucc}{cmyk}{1,0.4,0.2,0}
\definecolor{viola}{cmyk}{0,1,0,0.6}
\definecolor{viola2}{cmyk}{0,1,0.2,0.6}
\definecolor{verde}{cmyk}{0.92,0,0.59,0.25}
\definecolor{verdec}{cmyk}{0.92,0,0.59,0.15}
\definecolor{verdes}{cmyk}{0.92,0,0.59,0.4}
\font\tenrsfs=rsfs10 at 12pt
\font\sevenrsfs=rsfs7
\font\fiversfs=rsfs5
\newfam\rsfsfam

\textfont\rsfsfam=\tenrsfs
\scriptfont\rsfsfam=\sevenrsfs
\scriptscriptfont\rsfsfam=\fiversfs
\def\mathscr#1{{\fam\rsfsfam\relax#1}}

\newcommand{\vel}{{\cal V}}

\def\circa#1{\,\raise.3ex\hbox{$#1$\kern-.75em\lower1ex\hbox{$\sim$}}\,}

\usepackage{amsmath,latexsym,amssymb,color,hyperref,graphicx}
\newcommand{\eq}[1]{(\ref{#1})}
\newcommand{\be}{\begin{equation}}
\newcommand{\ee}{\end{equation}}
\newcommand{\bea}{\begin{eqnarray}}
\newcommand{\ena}{\end{eqnarray}}
\newcommand{\no}{\noindent}
\newcommand{\nb}{\nonumber}

\renewcommand\a{\alpha}
\renewcommand\b{\beta}

\newcommand\m{\ensuremath{\mu}}
\renewcommand\k{\ensuremath{\kappa}}

\newcommand\n{\ensuremath{\nu}}

\newcommand{\de}{\partial}

\newcommand{\ba}{\begin{eqnarray}}
\newcommand{\ea}{\end{eqnarray}}
\newcommand{\plm}{M_{\text{Pl}}}

\makeatletter
\def\ps@mine{%
    \def\@oddfoot{\hfil\thepage\hfil}\let\@evenfoot\@oddfoot
    \let\@oddhead\@evenhead%
    \let\@mkboth\@gobbletwo
    \let\sectionmark\@gobble
    \let\subsectionmark\@gobble
    }
\renewcommand\section{\@startsection {section}{1}{\z@}%
                                   {-3.5ex \@plus -1ex \@minus -.2ex}%
                                   {2ex \@plus.2ex}%
                                   {\normalfont\large\sffamily\bfseries}}
\renewcommand\subsection{\@startsection {subsection}{1}{\z@}%
                                   {-3.5ex \@plus -1ex \@minus -.2ex}%
                                   {2ex \@plus.2ex}%
                                   {\normalfont\sffamily\bfseries}}
\makeatother

\numberwithin{equation}{section}

\begin{document}
\thispagestyle{empty}
\vspace*{-2.5cm}
\begin{minipage}{.45\linewidth}

\end{minipage}
\vspace{2.5cm}

\begin{center}
{\huge\sffamily\bfseries 
Fluids, Superfluids and Supersolids:
Dynamics and Cosmology of Self-Gravitating Media
 }
 \end{center}
 
 \vspace{0.5cm}
 
 \begin{center} 
 {\sffamily\bfseries \large  Marco Celoria}$^{a}$,  
 {\sffamily\bfseries \large Denis Comelli}$^b$,
  {\sffamily\bfseries \large Luigi Pilo$^{c,d}$}\\[2ex]
  {\it
$^a$ Gran Sasso Science Institute (INFN)\\Via Francesco Crispi 7,
L'Aquila, I-67100\\\vspace{0.1cm}
$^b$INFN, Sezione di Ferrara,  I-35131 Ferrara, Italy\\\vspace{0.1cm}
$^c$Dipartimento di Fisica, Universit\`a di L'Aquila,  I-67010 L'Aquila, Italy\\\vspace{0.1cm}
$^d$INFN, Laboratori Nazionali del Gran Sasso, I-67010 Assergi, Italy\\\vspace{0.3cm}
{\tt marco.celoria@gssi.infn.it}, 
 {\tt comelli@fe.infn.it}, {\tt luigi.pilo@aquila.infn.it}
}
\end{center}

\vspace{0.7cm}

\begin{center}
{\small \today}
\end{center}

\vspace{0.7cm}

\begin{center}
{\bf \Large Abstract}
\end{center}

We compute  cosmological perturbations for a generic
self-gravitating media described by four derivatively-coupled  scalar fields.
Depending on the internal   symmetries  of the action for the scalar fields,
one can describe perfect fluids,  superfluids,  solids and supersolids media. 
Symmetries dictate both dynamical  and   thermodynamical properties of the media. 
Generically, scalar perturbations  include, besides the gravitational potential,  
an additional non-adiabatic mode associated with the entropy per particle  $\sigma$.
While perfect fluids and solids are adiabatic with $\sigma$  constant in time,
superfluids and supersolids feature a
non-trivial dynamics  for $\sigma$. Special classes of isentropic
media  with zero $\sigma$ can also be found.
Tensor modes  become massive for solids and supersolids.
Such an effective approach can be used to give a very general and
symmetry driven modelling of the  dark sector.

\no

\newpage
\section{Introduction}
An impressive amount of data  indicates that  the Universe is accelerating \cite{Planck-2015-par}
and a great effort is underway  to understand what is  driving such a phase~\cite{Kopp:2016mhm}.
Identifying the content of the dark sector is particularly challenging, thus
it is very useful to classify the various alternatives by using
symmetries. In our approach the dark sector is modelled as
a generic
self-gravitating medium with the only requirement to admit  an
isotropic Friedmann-Lemaitre-Robertson-Walker (FLRW) background
solution. In the hydrodynamical approximation, it turns out that a
generic   medium can be  effectively described by the theory of four
derivatively coupled scalar fields.  The four scalar fields can be
interpreted as comoving coordinates of the medium whose fluctuations
represent the Goldstone modes for the broken spacetime
  translations. The very same scalar fields can be viewed as
  St\"uckelberg fields that allow to restore broken 
diffeomorphisms~\cite{Leutwyler:1993gf,Leutwyler:1996er,ArkaniHamed:2002sp,ussgf,Rubakov:2008nh,Dubovsky:2004sg}.
Such an effective field theory description has been already considered
in~\cite{Dubovsky:2005xd,Dubovsky:2011sj,Nicolis:2011cs,Ballesteros:2012kv}
for particular type of media. From our analysis it turns out that the
internal symmetries of the medium action are crucial. Indeed, we will
show that dynamical and thermodynamical properties of the medium are determined by internal
symmetries which reflect on the form of the energy momentum tensor (EMT)
and give rise  to  conserved currents. Media can be
conveniently classified, according to the internal symmetries of the scalar
field theory~\cite{Dubovsky:2005xd,Nicolis:2013lma,Nicolis:2015sra,ussgf} in
perfect fluids, superfluids, solids and supersolids. In
  the present  unified approach, an important role is
played by the entropy per particle $\sigma$ in the dynamics of the
medium perturbations. 
Thermodynamical properties of a medium are studied by creating a
dictionary among the operators of the effective field theory and the
basic thermodynamical variables, extending to general media the
analysis of~\cite{usthermo}, see also \cite{Dubovsky:2011sj,Nicolis:2011cs}.

The main dynamical features of  linear cosmological
perturbations can be conveniently analysed by introducing a set of five
mass-like parameters $\{M_i \}$ related to first and second derivatives
of the Lagrangian density $U$ which resemble the masses used in
massive gravity
theories~\cite{Rubakov:2004eb,Dubovsky:2004ud,Rubakov:2008nh,Blas:2009my,Ballesteros:2012kv}. 
This is not a  coincidence: there is a close relationship between
massive gravity theories and the physics of self-gravitating media~\cite{Dubovsky:2005xd,ussgf}.
While the dynamical equations for cosmological perturbations are rather
cumbersome when expressed in terms of the fluctuations of
the scalar fields, they have a clear physical
interpretation once the entropy per particle $\sigma$ is
introduced. Generically, in the scalar sector, two dynamical modes
exist: the fluctuation of the gravitational potential and the perturbation $\delta\sigma$ of the entropy per particle.  
There are media, like perfect fluids, where the 
dynamics of $\delta \sigma$ is very simple: $\delta \sigma$ is conserved in time.
For superfluids and supersolids $\delta \sigma$ has a more complicated
evolution.

The outline of the paper is the following. In Section \ref{sect:EFT}
it is introduced the effective description of a generic rotational
invariant medium as the field theory of four derivatively coupled scalar fields.
In Section \ref{sect:thermo} we derive the correspondence between the
operators in the effective theory and the basic thermodynamical
variables of the medium. Section \ref{sect:cosmo}  is devoted to the study
of scalar, vector and tensor cosmological perturbations around a FLRW
spacetime. In Section \ref{DoFsec}   the number of
propagating degrees of freedom is related to the values of mass
parameters in the effective theory. 
In Section \ref{sect:special} we
characterise  adiabatic and  isentropic media. 
In Section \ref{sect:fluids} we study the class of 
Lagrangians that, at leading order, describes  perfect fluids.
 In Section \ref{sect:superfluids} the same analysis is carried out
 for superfluids and in Section \ref{sect:solids} for solids. In
 Section \ref{sect:supersolids} supersolids  are discussed, in particular we describe the
 dynamics of superhorizon scalar modes.   
 
 \section{Action Principle and Symmetries for Media Lagrangians}
\label{sect:EFT}
Non dissipative  media  can be described by using an effective
field theory  based on 
four St\"uckelberg scalar fields $\varphi^A$  ($A=0,1,2,3$), see for
instance~\cite{Dubovsky:2005xd,Dubovsky:2011sj,ussgf,usthermo} which can be related to
the Goldstone bosons for the spontaneous breaking of spacetime
translations~\cite{Leutwyler:1993gf,Leutwyler:1996er}. The medium physical
properties are encoded in a set of  symmetries of the scalar field
action  selecting, order by order in a derivative expansion, a finite
number of operators (see \cite{Endlich:2010hf,Dubovsky:2011sj, Ballesteros:2014sxa, Gripaios:2014yha} for a next to leading study of the perfect fluid case). At the leading order the fundamental object is
\be
C^{AB}= g^{\mu \nu} \de_\mu \varphi^A \de_\nu \varphi^B  \, ;
\ee
where $g_{\mu \nu}$ is the spacetime metric. The effective medium
action is built assuming diff invariance and internal rotational
invariance, namely $\varphi^a \to R^a_b \, \varphi^b \, , \; a,b=1,2,3;$
with $\pmb{R} \in SO(3)$.   In what follows, we will always use boldface capital letters for three-dimensional spatial matrices.
Since the fields $\varphi^a$ can be interpreted as comoving
coordinates, this allows to define a (unique) four-velocity $u^\mu$, through the conditions \cite{Carter:1987qr}
\be
u^\mu \, \de_\mu \varphi^a =0 \, , \qquad u^\mu u^\nu g_{\mu \nu} =-1\,,
\label{fvel}
\ee
whose only solution is
\be
u^\mu = -\frac{\epsilon^{\mu \nu \alpha \beta}}{6 \, b\sqrt{-g} }\epsilon_{abc}
\, \de_\nu \varphi^a \, \, \de_\alpha \varphi^b \, \de_\beta \varphi^c \,
 \,,
\label{veloc}
\ee
where
\be \label{detb}
\qquad b \equiv\sqrt{\det\,\pmb B}\,,
\ee
and $\pmb B$ denotes the $3\times 3$ matrix whose components are
$B^{ab} \equiv C^{ab}$.
When pulled-back into the spacetime, the medium metric $B_{ab}$ becomes a projector $h_{\mu \nu}$
\be 
h_{\mu \nu} = B_{ab} \de_\mu \varphi^a \, \de_\nu \varphi^b
\equiv g_{\mu \nu} + u_\mu \, u_\nu \, , \qquad h_{\mu \nu} u^\nu=0 \, ,
\label{ind}
\ee
and $B_{ab}$ indicates the matrix elements of $\pmb {B}^{-1}$.
Moreover, we impose the condition
\begin{equation}
X=C^{00}<0\, 
\end{equation}
which allows to define another  time-like four-velocity 
\be 
\label{vdef} 
\vel^\mu = -\frac{\partial^\mu\varphi^0}{\sqrt{{-X}}} \, , \qquad \qquad\vel^2 =-1 \,.
\ee

Following~\cite{ussgf}, 
the operators with definite transformation properties under
internal rotation  built from  $C^{AB}$ are listed in Table \ref{tab:operators}.
\begin{table}[h!]
  \footnotesize
  \small
  \renewcommand\arraystretch{1.5}
  \centering
  \begin{tabular}{|c|c|}
    \hline{\bf Operator} &{\bf Definition}    \\ 
    \hline
    \hline
$C^{AB}$ & $g^{\mu \nu} \, \de_\mu \varphi^A \, \de_\nu \varphi^B$\,,\quad \scriptsize{$A,B=0,1,2,3$}\\  \hline 
$B^{a b}$ & $g^{\mu \nu}\partial_\mu \varphi^a \, \partial_\mu \varphi^b$\,,\quad \scriptsize{$a,b=1,2,3$}\\ \hline 
$Z^{ab}$ & $C^{a0} \, C^{b0}$ \\ \hline
$X$  & $C^{00}$\\ \hline
$W^{ab}$ & $B^{ab} - Z^{ab}/X$\\ \hline
 $b$ &  $\sqrt{\det \pmb{B}}$ \\ \hline
  $Y$    & $u^\mu \partial_\mu \varphi^0$ \\ \hline
\end{tabular}
\hskip 1cm
\begin{tabular}{|c|c|}
 \hline{\bf Operator} &{\bf Definition}    \\ 
    \hline
    \hline
   $y_n$ & $\text{Tr} \left({\pmb B}^n \cdot \pmb{Z} \right)$\,, \quad \scriptsize{$n=0,1,2,3$}\\ \hline
$\tau_n$ & $\text{Tr}\left(\pmb{B}^n \right)$\,, \quad \scriptsize{$n=1,2,3$}\\ \hline
$w_n$ & $\text{Tr}\left(\pmb{W}^n \right)$\,, \quad \scriptsize{$n=0,1,2,3$}\\ \hline
${\cal O}_{\alpha\beta n}$ &  $\left({X}/{Y^2}\right)^\alpha \, \left({y_n}/{Y^2}\right)^\beta$\,,\quad \scriptsize{$\alpha\,,\beta \in  {\bf R} $} \\ \hline
${\cal O}_{\alpha}$ &  $\left({X}/{Y^2}\right)^\alpha$\,,\quad
                      \scriptsize{$\alpha \in {\bf R} $}  
\\ \hline
\end{tabular}
  \caption{\it Operators summary. Greek letters for spacetime indices, capital Latin letters  {\scriptsize ($A,B,\ldots=0,1,2,3$)}  for indices in the internal spacetime of the medium and small Latin letters {\scriptsize ($a,b,\ldots=1,2,3$)} for spatial indices of the medium \cite{ussgf}.}
  \label{tab:operators}
\end{table}

\no
The most general action at LO for a medium described by the four $\varphi^A$ can be constructed in
terms of the scalar invariants $X$, $Y$, $\tau_n$ and $y_n$ with $n=1,2,3$ and $m=0,..,3$ for a total of nine independent operators
\cite{ussgf}
 \be
\label{LAll}
 S=\plm^2\;\int d^4 x\;\sqrt{-g}\, R+\int d^4x\sqrt{-g} \; U(X, \,
 Y, \,  \tau_n, \, y_n) \,.
 \ee
By imposing further symmetries besides internal rotational invariance,
media can be characterised  according to Table \ref{tab:summa}. For a
detailed analysis see~\cite{ussgf}.
\begin{table}[h!]
  \footnotesize
  \small
  \renewcommand\arraystretch{1.5}
  \centering
  \begin{tabular}{|c|c|c|}
 \hline
   \multicolumn{3}{|c|}{\bf Four-dimensional media}\\
    \hline
    \hline
{\bf Symmetries of the action} &{\bf LO scalar operators} &  {\bf Type of medium}    \\ 
    \hline
    \hline
$SO(3)_s$\quad \&\quad $\varphi^A\to\varphi^A+f^A$\,,\quad $\partial_\mu f^A=0$ &   $X$, $Y$, $\tau_n$, $y_n$ &  
{supersolids}  \\ \cline{1-2}
  $\varphi^a\to \varphi^a+f^a(\varphi^0)$   &  $X$, $w_n$ & \\ \cline{1-2}
$\varphi^0\to \varphi^0+f(\varphi^0)$   & $\tau_n$, $w_n$, $\mathcal{O}_{\alpha\beta n}$ &  \\ \cline{1-2}
  $\varphi^a\to \varphi^a+f^a(\varphi^0)$\, \&\, $\varphi^0\to \varphi^0+f(\varphi^0)$  &  $w_n$ &  \\ \hline 
   $\varphi^0\to \varphi^0+f(\varphi^a)$   & $Y$, $\tau_n$ & {solids} \\\cline{1-2} \hline
$V_s$Diff: $\varphi^a\to\Psi^a(\varphi^b)$\,,\quad$\det |\partial \Psi^a /\partial{\varphi^b}|=1$  &  $b$, $Y$, $X$ & 
{superfluids} \\\cline{1-2}
        $\varphi^0\to \varphi^0+f(\varphi^0)$ \, \&\, $V_s$Diff  &  $b$, $\mathcal{O}_{\alpha}$ &  \\ \hline
    $\varphi^0\to \varphi^0+f(\varphi^a)$ \, \&\, $V_s$Diff  &  $b$, $Y$ & perfect fluid \\ \hline
$\varphi^A\to\Psi^A(\varphi^B)$\,,\quad$\det |\partial \Psi^A 
       /\partial{\varphi^B}|=1$  &  $b\, Y$ &  perfect fluid with $\rho+p=0$ \\\hline 
       \end{tabular}
  \caption{\it Summary of local symmetries in material spacetime and
    the corresponding invariant scalar operators. Invariance under $SO(3)_s$ and shift symmetries are assumed by default in all cases \cite{ussgf}.}
  \label{tab:summa}
\end{table}
One can also consider media  with reduced internal dimensionality for
which only the fields $\varphi^a$ are relevant. For instance, $U(b)$
describes a perfect fluid while $U(\tau_1, \tau_2 , \tau_2)$ is a solid.

\section{Conserved Currents, EMT and Thermodynamics}
\label{sect:thermo}
As a consequence of the shift symmetry  $\varphi^A \rightarrow \varphi^A +
c^A$, the equations of motion derived from (\ref{LAll}) have the form
\begin{equation}
2 \, \nabla_\mu (U_{C^{AB}} \partial^\mu \varphi^B) = 0 = \nabla_\mu J^{\mu A}
\end{equation}
where the $J^\mu_A$ are the Noether currents for the shift symmetry 
\begin{equation}
J^\mu_A=
\frac{\partial U}{\partial ( \partial_\mu \varphi^A)}=2\;U_{AB} \;\nabla^\mu \varphi^B \, , \qquad \quad U_{AB} = \frac{\partial
  U}{\partial C^{AB}} \, ;
\end{equation}
the expression of $J^\mu_A$ are given in appendix \ref{emta}. 
The energy-momentum tensor (EMT) derived from (\ref{LAll}) 
(see appendix \ref{emta} for details) can be written in terms of the
Noether currents as
\begin{equation}
T_{\mu\nu}= - 2\;\frac{\delta (\sqrt{-g}\;U)}{\sqrt{-g}\;\delta g^{\m\n}}=
U\; g_{\mu\nu} -J_{A \mu} \;\partial_\nu\varphi^A \, .
\label{emtgen}
\end{equation}
The conservation of the EMT, $\nabla_\nu T^{\mu \nu} =0$, is
equivalent to equations of motion of the scalar fields.
It is convenient to define the following tetrad $\{
\beta^A_\mu \, , A=0,1,2,3\}$ defined by
\be
\beta^0_\mu= u_\mu \,  , \qquad h_{\mu \nu} = \beta_\mu^a \,
\beta^b_\nu \;\delta_{ab} \, , \quad \text{with } \beta_\mu^A \,
\beta_\nu^B\; g^{\mu \nu} = \eta^{AB} \, .
\ee
and the vector
\begin{equation}
\xi_\mu=-h_{\mu\nu}\; \partial^\nu \varphi^0=-Y\;u_\m +
\sqrt{-X}\;\mathcal{V}_\m
\label{csi}
\end{equation} 
which is related to the   relative velocity of the superfluid/supersolid
component with respect to the normal 
component~\cite{Nicolis:2011cs, Landau_Fluid_Mechanics}.
Projecting the EMT (\ref{emtgen}) on $\{\beta^A_\mu \}$, we have
\be
 \begin{split} 
&T_{\mu \nu} = \rho \, u_{\mu} \, u_{\nu} + q_\mu \, u_\nu + q_\nu \,
u_\mu+ {\cal P}_{\mu \nu} \, ;\\
& \rho = T^{\mu \nu}\, u_\mu \,u_\nu \, , \qquad q_\mu =- h_{\mu \alpha}\,
T^{\alpha \beta} \,u_\beta \, , \qquad {\cal P}_{\mu \nu} = h_{\mu
 \alpha} \,h_{\nu \beta}\, T^{\alpha \beta} \, .
\end{split}
\label{sgmemt}
\ee
The EMT formally has the form of imperfect fluid with heat flow 
\be
q_\mu = 2\;Y \left[   \, \sum_{m=0}^3 U_{y_m}
  \left(\boldsymbol{B}^m \right)^{ab} \, \nabla_\mu \varphi^b \, C^{a0}
-  \, U_X \, \xi_\mu \right] \, .
\ee
Writing the current $J_0^\mu$  as
\be
J^\mu_0 = \left(U_Y - \, 2 Y \, U_X \right) u^\mu + \frac{q^\mu}{Y} 
\ee
it formally coincides with the entropy current $s^\mu$ of an imperfect
fluid~\cite{rezzolla2013relativistic}
\be
s^\mu = s \, u^\mu + \frac{q^\mu}{T} \, .
\ee
Thus, it is natural to identify the temperature $T$ with $Y$ and the
entropy density $s$ with $U_Y - \, 2 \,Y \, U_X $. Notice that of course
no dissipation is present being $\nabla_\mu J^\mu_0=\nabla_\mu s^\mu=0$. 
The current 
\be
J_\mu=b\;u_\mu \, ,\qquad \nabla^\mu J_\mu =0 
 \label{cons}
\ee
is conserved off shell  and can be identified with the particle density
current. Splitting the tensor ${\cal P}_{\mu \nu}$ in a trace and a traceless part 
\be\label{Tmn2}
{\cal P}_{\mu \nu}= h_{\mu \nu} {\cal P} + {\cal P}_{\mu
  \nu}^{\text{tl}} \, , \qquad  {\cal P} = \frac{ {\cal P}_{\mu \nu}
    h^{\mu \nu}}{3} \, , \qquad  {\cal P}_{\mu
  \nu}^{\text{tl}} h^{\mu \nu} =0 \ ;
\ee
we have that (see eq.(\ref{sigma0}-\ref{sigmaa}))
\be
{\cal P} = p + \Pi \, , \qquad p = U - \frac{1}{3} \, J^\alpha_a
\, h^\mu_\alpha \, \nabla_\mu \varphi^a  \, , \qquad \Pi = \frac{q_\mu \xi^\mu}{3 \, Y} \,.
\ee
The ``perfect'' part $p$ of ${\cal P}$ is identified with the thermodynamical
pressure, while the viscosity $\Pi$ is a component  of the anisotropic stress.  
As a result 
\ba
\label{rhop}
 && \rho= -U+Y\;U_Y-2\;Y^2\; U_X \, ;\\  
 && {\cal P}= U-b\;U_b-\frac{2}{3}\;\sum_{n=1}^3\;n\;\tau_n\;U_{\tau_n}
-\frac{2}{3}\,\sum_{n=0}^3\;n\;y_n\;U_{y_n}-\frac{2}{3}\;U_X\;(Y^2+X)\\  
&& p=U-b\;U_b-\frac{2}{3}\;\sum_{n=1}^3\;n\;\tau_n\;U_{\tau_n}
-\frac{2}{3}\,\sum_{n=0}^3\;(n+1)\;y_n\;U_{y_n}
 \, .  \label{rhop1}
 \ea
A complete treatment of the thermodynamics of a self-gravitating
medium will be given elsewhere. Here we simply outline general idea. Extending the very same reasoning in~\cite{usthermo} to general
media, the particle number density $n$ and the entropy density $s$ can
be associated to the following projections of the conserved currents in the $u^\m$ frame
\ba
n=-J_\a\;u^\a=b,\qquad s=-J^\alpha_0 \, u_\a=U_Y-2\;Y\;U_X \, .
\label{charges}
\ea
An important quantity in the dynamics of self gravitating media is the
entropy per particle $\sigma=s/n$, whose evolution can be obtained from the conservation of $J_\mu$
and $s_\mu$
\ba
\label{sig}
u^\a\nabla_\a
\sigma=-\frac{2}{b}\;\nabla^\a\;\left[-
  \;U_X\; \xi_\alpha 
  +\;\sum_{n=0}^3 \; U_{y_n}\;
   {\cal C}^{0a} \; ({B}^n)^{ab} \; \partial_\a  {\varphi}^b\right] \, =-\frac{1}{b}\;\nabla^\a\;\left(\frac{q_\alpha}{Y}\right).
\ea
 While for perfect fluids, described by $U(b,Y)$, $\sigma$ is conserved as
 expected,  for media in which the operators $X$ and
$y_n$ are relevant (for instance  superfluids and supersolids)  
$\sigma$ is not conserved.

\no
The presence of a superfluid/supersolid component requires  at least
an additional thermodynamical variable \cite{khalatnikov1982relativistic,PhysRevD.45.4536,Nicolis:2011cs,Comer:2011ss,Andersson:2006nr}
$\xi$, related to the space like vector
$\xi_\mu$, see eq.(\ref{csi}), namely
\be
\xi= \sqrt{\xi_\m\xi^\m}  =\sqrt{X+Y^2}\, .
\ee
Indeed, by extending  the reasoning
in~\cite{usthermo} one can use as fundamental thermodynamical
variables  the densities $n$, $s$, any function $f(\xi)$ of $\xi$ and the conjugate variables  $\mu$, $T$ and $\eta$~\cite{Landau_Fluid_Mechanics,Callen}. 
Setting for simplicity $f(\xi)=\xi$, the thermodynamical dictionary is obtained by taking 
the operators $b$, $Y$, $X$, $\tau_i$ and $y_n$  as functions of the
thermodynamical variables such that the first principle
\begin{equation}
d\rho=T\;ds+\mu\;dn + \eta \; d\xi \, ,
\label{first}
\end{equation}
and the Euler relation in term of the thermodynamical pressure $p$
\begin{equation}
\rho+p=T\;s+\mu\, n  \, 
\label{euler}
\end{equation}
are satisfied.
One can verify that (\ref{first}-\ref{euler}) hold when
\be
\begin{split}
&Y =T \, , \qquad s = U_Y - \, 2 \;Y \, U_X \, , \qquad  b=n  \, , \\
&\xi = (X+Y^2)^{1/2}  \, , \qquad \eta = -2  \, U_X
\, \xi -2 \, \xi^{-1} \sum_{m=0}^3 y_m \, U_{y_m} \; \\
& \mu = -U_b -\frac{2}{3 \, b}\;\left[ \sum_{m=1}^3 m \, \tau_m \, U_{\tau_m}
  +\sum_{m=0}^3 (m+1) \, y_m \, U_{y_m} \right] \, , \qquad \tau_m =
n^{2m/3} \, , \qquad y_m = n^{2(m+1)/3} \, \xi^2 \, .
\label{ch1}
\end{split}
\ee
Thermodynamics of fluids and superfluids was considered also in
\cite{Dubovsky:2011sj},\cite{Nicolis:2011cs},\cite{Nicolis:2013lma} and 
 our results  for such 
subcases are in agreement, except for  exchange of $s$ with $n$ (and the corresponding conjugate variables $T$ and $\mu$).
Notice that the Euler relation (\ref{euler}) is somehow peculiar being
independent from $\xi$ while this is not the case for  the first
principle. Introducing the
generalized pressure  $P$ by a Legendre transform of
 the  thermodynamical pressure $p$ in eq. (\ref{rhop1}) with respect to
 $\xi$, namely 
\ba
P=p+\eta\;\xi=U-b\;U_b-\frac{2}{3}\;\sum_n\;n\;\tau_n\;U_{\tau_n}
 -2\;(X+Y^2)\;U_X-\frac{2}{3}\,\sum_{n=0}^3\;(n+4)\;y_n\;U_{y_n} \, ;
\ea  
we have that the following improved Euler relation holds
\be
\rho + P = T \, s + n \, \mu+ \xi \, \eta \, .
\ee
%

As far as thermodynamics is concerned, all operators of the EFT can be
considered as function of $\rho$, $\sigma$ and $\xi$, thus the 1-form $dp$ can  be written in terms of the 1-forms $d\rho$, $d\sigma$ and $d\xi$, namely
\be
\label{dp}
 dp  =
 \left.\frac{\partial p}{\partial \rho}\right|_{\sigma, \xi}\;d\rho
 +\left.\frac{\partial p}{\partial \sigma}\right|_{\rho,\xi}\; d \sigma
 +\left.\frac{\partial p}{\partial \xi}\right|_{\rho,\sigma}\; d\xi
\ee
whose coefficients will be explicitly  computed  in Section \ref{sect:cosmo} at leading order  in  a FLRW spacetime (being $\xi  =y_n =0$ on FRW, no difference at such order is present in between $p$ and $P$).

\section{Cosmology of Self Gravitating Media}
\label{sect:cosmo}
In this section we will study isotropic cosmological solutions and their 
linear perturbation dynamics.
The Einstein equations are of the form
\be
R_{\mu \nu} -\frac{R}{2} \, g_{\mu \nu} = \frac{ T_{\mu
  \nu} }{2 \, \plm^2}\, ;
\ee
where the EMT is given  by (\ref{sgmemt}). 
We consider the spatially flat Friedmann-Lemaitre-Robertson-Walker FLRW
background 
\be
ds^2 = a(t)^2 \; \eta_{\mu \nu} \;dx^\mu\; dx^\nu   \, ;
\ee
where conformal time has been used and $\eta_{\m\n}$ is the Minkowski
metric. The  scalar fields $\varphi^A$ assume the following background
value form compatible with isotropy and homogeneity
\be
\bar \varphi^0=\phi(t),\qquad \bar \varphi^a=x^a  \, .
\label{sbkg}
\ee

It is very convenient to introduce a number of mass parameters defined in
the so called unitary gauge~\footnote{In the Standard Model, the  tree level
masses of gauge bosons related to the breaking of $SU(2)\times U(1)$
are defined in the unitary  gauge where the would-be Goldstone bosons
are set to zero.} where  the scalar fields
fluctuations are gauged away and then their value coincides with the
background form  (\ref{sbkg})
\be
\varphi^A_{(ug)} = \bar \varphi^A \, .
\ee
Thus, all perturbations  are  pure metric perturbations
\be
 g_{\mu\nu}^{(ug)}(t,\,\vec x)=a^2 \, \left(\eta_{\mu\nu}+h_{\mu\nu}^{(ug)}(t,\,\vec x)
 \right) \, .
\ee
The  Lagrangian  $U$ in the action \eq{LAll} can be expanded up to second order  as follows \cite{Rubakov:2004eb,Dubovsky:2004sg,Dubovsky:2004ud,Rubakov:2008nh,Blas:2009my}
\ba\nonumber
\sqrt{- g} \, U 
&\equiv&  \frac{\sqrt{-\bar{g}}}{2}\;\bar{T}^{\mu \nu} \;a^2\;
  \textsl{h}_{\mu \nu}^{(ug)} 
  + \frac{\plm^2} {4} \left[\lambda_0^2 \;  h_{00}^{(ug)}{}^2+2\,\lambda_1^2\;  h_{0i}^{(ug)\,2}
    -2\,\lambda_4^2 \; h_{00}^{(ug)}\,  h_{ii}^{(ug)} +\lambda_3^2 
\;  h_{ii}^{(ug)}{}^2  - \lambda_2^2\;   h_{ij}^{(ug)\,2} 
\right];
\\\label{masst}
 && \bar{T}^{\mu\nu}= \frac{1}{a^2}\;(\bar{\rho}\;\delta^\mu_0\;\delta^\nu_0\;  +\bar
 p\;\delta^\mu_i\;\delta_{ij}\;\delta^\nu_j ) \, .
  \ea
Where $\bar{\rho}$ and $\bar{p}$ are the background values for the
energy density and pressure. From the rotational invariance of the background, we see that the
quadratic expansion of $U$ is fixed by giving seven time dependent parameters:
$\bar\rho,\;\bar p,\;\lambda_{0,1,2,3,4}$. In particular
\be
\begin{split}
& \bar{\rho} =\frac{U_Y \;\phi'}{a} -\frac{ 2 \;U_X \, \phi '^2}{a^2}-U \, ;\\
& \bar{p} = U-\frac{6 \;U_{\tau_3}}{a^6}-
\frac{4\; U_{\tau_2}}{a^4}-\frac{U_b}{a^3}-\frac{2 \;U_{\tau _1}}{a^2} \, ;
\end{split}
\label{bkg}
\ee
 It is understood that $U$ and its derivatives are computed at the
background values of EFT operators which read 
\ba
\bar\vel^{\mu}= \bar u^{\mu}=\left(\frac{1}{a} \, ,\;\vec{0}\right) \,
,\quad
\bar b=\frac{1}{a^3} \, ,\quad \bar Y=\frac{\phi'}{a} \, ,\quad \bar
X=-\left(\frac{\phi'}{a}\right)^2 \, ,\quad
\bar\tau_n=\frac{3}{a^{2\,n}} \, ,\qquad \bar y_n=0 \, .
\ea
Notice  that at background level we have $\bar\xi=\sqrt{\bar X+\bar Y^2}=0$.

 It is convenient to introduce the following combinations that will be
 instrumental to study the dynamics of perturbations
 \ba
\label{MM}
 &&M_0= \lambda_0^2\;-\frac{a^4\;\bar{\rho}
   }{2\;\plm^2},\qquad   M_{1}=\lambda_{1 }^2 - a^4\; \frac{\bar p}{\plm^2} ,\qquad
   M_{2 }=\lambda_{2 }^2 -a^4\; \frac{\bar p}{\plm^2} \, ,
   \\&&\nonumber
   M_{3 }=\lambda_{3 }^2 -\frac{a^4\; \bar p}{2\;\plm^2},\qquad
   M_{4 }=\lambda_{4 }^2-\frac{a^4\; \bar p}{2\;\plm^2}  \, .
  \ea
The  parameters $\{M_i\}$ have dimension 2 and their explicit value is given in appendix \ref{massesapp}. 
When the Lagrangian $U$ has a global Lorentz invariance around Minkowski (where $\phi'=1$), at least at
quadratic level, the masses $\lambda_i$  are more constrained and can be
expressed in terms of only two parameters $A$ and $B$ as 
$\lambda_0^2=A+B$ and $\lambda_{1,2}^2=-A$,  $\lambda_{3,4}^2=B$.
The conservation of the background EMT is equivalent to the 
equation of motion for $\bar \varphi^0 = \phi$
\be
\label{eqphi}
\bar{\rho}'=-3\;{\cal H}\;(\bar{\rho}+\bar{p})\quad\Rightarrow \quad
   \phi ''=  \left(1-3\;
   c_b^2\right)\;\mathcal{H}\; \phi ' \, .
 \ee
From the definitions \eq{MM} we can define two parameters $c_{b }^2$ and $c_{ s}^2$ whose precise meaning and relation with thermodynamics will be given later (see Eq. (\ref{dpTb}) for $c_s^2$ and Eq. (\ref{Gamma}) for $c_b^2$ )
\be
   c_b^2= -\frac{M_4}{M_0} \, ,\qquad 
c_s^2=\frac{2 \;\plm^2\;\left[3\, M_4^2+M_0\,
   \left(M_2-3 \,M_3\right)\right]}{3\, a^4\, M_0 \,(\bar{p}+\bar{\rho})} \, .
\label{backg}
\ee
The background Einstein equations determine the 
evolution of the scale factor $a$
\be
\mathcal{H}^2 \equiv
\left(\frac{a'}{a}\right)^2=\;\frac{a^2\;\bar{\rho}}{6\, \plm^2} \, ,\qquad
\mathcal{H}'=-a^2\;\frac{(\bar{\rho}+3\;\bar{p})}{12\, \plm^2} \, .
\ee
From the conservation of the currents \eq{cons} and from \eq{sig}, evaluated at
at background level, it follows that 

\be
\label{sig0}
\begin{split}
& \bar s'+3\;{\cal H}\;\bar s=0,\quad \bar n'+3\;{\cal H}\;\bar n=0\,
;\\
& \bar \sigma\equiv\frac{\bar s}{\bar n},\quad \bar\sigma'=0 \, .
\end{split}
\ee
Thus, the  entropy per particle is always conserved on a FLRW
background as a consequence of the perfect fluid form of the
background EMT.
Defining  the equation of state parameter $w= \bar{p}/\bar{\rho}$, we
have that 
\be
\label{pG}
  w'=-3\;{\cal H} \;(1+w)\;(c_s^2-w) \, ;
 \ee
where  $c_s^2 = \bar{p}\,'/ \bar{\rho}\,'$ is precisely the adiabatic sound
speed. Notice that  when $w$ is constant in time, then $c_s^2 =w$.
In the following we will first study the dynamics of perturbations for generic values
of $M_i$. Typically, the vanishing of some of the $M_i$ corresponds to
an enhanced internal symmetry  as summarized in Table \ref{tab:femt}
where the mechanical properties (the structure of the EMT) are related
to special values of  $\{M_i\}$,  see (\ref{MXY}).

\begin{table}[h!]
  \footnotesize
  \small
  \renewcommand\arraystretch{2.2}
  \centering
  \begin{tabular}{|c|c|c|c| }
    \hline
 EMT & Lagrangian & Medium & Masses \\
    \hline \hline
    $q_\m=0,\;{\cal P}^{tl}_{\m\n}=0$                            & $U(b,\;Y)$ &                Perfect Fluid & $M_{1,\,2}=0$    \\ \hline
    $q_\m=0,\;{\cal P}^{tl}_{\m\n}\neq0$                        & $U(b,\,\tau_n,\,Y)$&   Solid            &  $M_1=0$ \\ \hline
    $q_\m\neq 0,\;{\cal P}^{tl}_{\m\n}\propto\;q_\m\;q_\n$ & $U(b,\,Y,\,X)$ & SuperFluid          &$M_2=0$    \\ \hline
    $q_\m\neq 0,\;{\cal P}^{tl}_{\m\n}\neq 0$                  &$U(b,\,Y,\,X,\,\tau_n,\,y_n)$ & SuperSolid  &$M_{1,2} \neq 0$  \\ \hline
\end{tabular}
 \caption{\it \small Media classification according to the EMT tensor
   properties and the relation with the values of $M_i$. 
 \label{tab:femt}}
\end{table}
 
\subsection{Scalar Perturbations}
In the Newtonian gauge,  at  linear order, the perturbed metric in the scalar
sector is given by
\be
ds^2=a^2 \, \eta_{\mu \nu}\, dx^\mu dx^\nu + 2 \, a^2 \left [ \Psi(t,\,\vec x) \, dt^2+
\Phi(t,\,\vec x) \, d\vec{x}^2 \right ] \, ;
\ee
where $\Phi$, $\Psi$ are the two Bardeen potentials \cite{Bardeen:1983qw,Kodama:1985bj}.
 At the background level,  the EMT of the medium in a homogeneous and isotropic universe has to take the form of a perfect fluid. 
A generic  perturbed perfect fluid EMT can be written as
\be
T_\nu^\mu=(\bar{p}+\bar{\rho}+\delta p+\delta\rho)\;\bar{\mathfrak{U}}^\mu\;\bar{\mathfrak{U}}_\nu+(\bar{p}+\bar{\rho})\;(\delta
\mathfrak{U}^\mu\;\bar{\mathfrak{U}}_\nu+\bar{\mathfrak{U}}^\mu\;\delta
\mathfrak{U}_\nu)+ (\bar{p}+\delta p)\;\delta^\mu_\nu+\Pi^\mu_\nu \, ,
\ee
where $\bar{\mathfrak{U}}_\mu=(-a,\vec{0})$ is the background 4-velocity, $\delta \rho,\;\delta p$ are the perturbations of energy density and
pressure. In the scalar sector, the  velocity $
\delta\mathfrak{U}^\m $  and the anisotropic stress perturbations  $\Pi^\mu_\nu$ 
can be written in terms of two extra scalars $v$ and $\Xi$ defined as\footnote{Note that the dimension of these two extra scalars is $[v]=-1$ and $[\Xi]=2$.}
\be
\delta \mathfrak{U}_\mu=(a\;\Psi,\;   a \, \partial_i  v),\qquad
\Pi^\mu_\nu\equiv\;
(3\;\partial^2  \, \delta^\mu_i\; \delta^i_\nu - \delta^\mu_i
\, \partial^i \, \delta_\nu^j \partial_j)\; \Xi \, .
\ee
With respect to equations \eq{sgmemt} and \eq{Tmn2} we can do the matching 
\ba\rho=\bar{\rho}+\delta \rho,\qquad p+\Pi=\bar{p}+\delta p,\qquad
q_\mu=(\bar{p}+\bar{\rho})\;\delta\mathfrak{U}_\m,\qquad {\cal P}^{tl}_{\m\n}=\Pi_{\m\n} 
\ea

In Fourier space, the linear perturbed Einstein equations then read
\ba
\label{eqEE}
a^2\;\delta\rho \!\!&=& \!\! 4\;\plm^2\;\left[ k^2\;\Phi+3\;{\cal H}\;(\Phi'+{\cal H}\;\Psi)\right]\\\nonumber
6 a^2 (w+1) {\cal{H}}^2 \, \pi _l'+M_1 \left(\pi _l'-\frac{1}{\phi'}\right)&=&\!\! 
2\; \left(\Phi'+{\cal H}\;\Psi\right)
\\ \nonumber
 \Xi  \!\!&=& \!\! 2\;\plm^2\; \left(\Phi- \Psi\right)\\\nonumber
a^2\;\delta p  \!\!&=& \!\! \!\! -\frac{4\;\plm^2}{3}\left
  \{k^2\;\Phi -\Psi\; \left[9\; w\;\mathcal{H}^2 +k^2\right]+
  3\;{\cal
    H}\;(\Psi'+2\;\Phi')+3\;\Phi''\right \}\nonumber
\ea
where $k^2 = k^i k^j \delta_{ij}$, with $k^i$ is the comoving momentum.

At background level, by using the fact that $\bar \xi =0$ and  $ \bar \sigma'=0$,
Eq. \eq{dp}  gives
\be\label{dpTb}
  {\bar{p}}\,'\equiv \left.\frac{\partial  {\bar p}}{\partial {\bar{\rho}}}\right|_{\bar\sigma}\;  {\bar{\rho}}\,'+
 \left.\frac{\partial  {\bar{p}}}{\partial
     {\bar\sigma}}\right|_{\bar{\rho}}\;  {\bar \sigma}\,' =c_s^2 \, 
 {\bar{\rho}}\,' \;,
 \qquad \qquad
 \left.\frac{\partial  {\bar{p}}}{\partial
     {\bar{\rho}}}\right|_{\bar\sigma}=\frac{{\bar p}\,'}{{\bar{\rho}}\,'}=c_s^2 \, .
 \ee
Notice that, in the expansion  of \eq{dp} at linear
  order, no contribution from  $\xi$  is present:
 $\delta\xi$ is of order two (see Eq. \eq{byx})
 being all spatial velocities zero on a FLRW by isotropy. This allows us to
write the total pressure variation as an adiabatic contribution
proportional to the energy density perturbation $\delta \rho$ and a non adiabatic part
proportional to the entropy per particle  perturbation $\delta \sigma$:
\be\label{dpp}
 \delta p\equiv c_s^2\;\delta \rho+ \left.\frac{\partial  {\bar p}}{\partial {\bar\sigma}}\right|_{\bar{\rho}}\;  \delta{  \sigma}
   \equiv c_s^2\;\delta \rho+\Gamma\, , \qquad \qquad   \Gamma = \left.\frac{\partial  {\bar p}}{\partial {\bar\sigma}}\right|_{\bar{\rho}}\;  \delta{  \sigma}  \, .
 \ee
For barotropic fluids with $\bar p=\bar p(\bar{\rho})$, we
have that $\frac{\partial  {\bar p}}{\partial {\bar\sigma}}=0$. As a
consequence,  $\Gamma$ is zero also when $\delta\sigma\neq 0$,
i.e. the entropic fluctuations do not  back react on the
evolution of the Bardeen potentials. 
Combining (\ref{eqEE}) and (\ref{dp}), one gets a second
order  ODE for the Bardeen potential $\Phi$
\be
\label{me}
\Phi '' +3 \, \mathcal{H}\,
   \left(\mathit{c}_{\mathit{s
   }}^2+1\right) \Phi '+\Phi\, 
   \left[k^2\,
   c_s^2+3\, \mathcal{H}^2 \left(c_s^2-  w\right)\right]=
-\frac{\hat\Xi}{6} \, \left[
   k^2- 9\, \mathcal{H}^2
   \left(c_s^2- w\right)\right]
 +\frac {\mathcal{H}\, \hat \Xi'}{2} - \frac{a^2 \, \hat \Gamma}{4} \, ;
\ee
we have defined $\hat \Xi=  \plm^{-2} \, \Xi$ and $\hat \Gamma=  \plm^{-2} \, \Gamma$. 

\no
Given $\Gamma$ and $\Xi$, the perturbed Einstein equations allow to
determine $\Phi$ through Eq. \eq{me} and then in turn $\delta \rho$ and $v$ \eq{eqEE}.
Actually, the EFT formulation and its thermodynamical interpretation
determine  uniquely both $\Gamma$ and $\Xi$. Indeed,
given $U$, the masses $M_i$ are defined and  $\delta \rho$,  $\delta
p$ and $v$ can be expressed in terms of scalar fields $\varphi^A$ and the
Bardeen scalars. At   linear level, by using the
rotational invariance of the background, the scalar part
of $\varphi^A$  can be written as
\be
\varphi^0=\phi(t)+\pi_0(t,\vec{x}),\qquad \varphi^a=x^a+\partial_a
\pi_L(t,\vec{x}) \, .
\ee
From the expressions \eq{rhop}, \eq{sig} we get
\be
\label{eqPP}
\begin{split}
 &
   {\delta \rho}=- \frac{6\, \plm^2\,\mathcal{H}^2}{a^2}\left( 3\, \Phi+k^2 \,\pi _L \right) \;
   (1+w)+\frac{\phi'}{a^4}\;\delta\sigma \, ;\\ 
 &\delta p =-  \frac{6\, \plm^2\,\mathcal{H}^2}{a^2}\,c_s^2\;\left( 3\, \Phi+k^2 \,\pi _L \right) \;
(1+w)+\frac{c_b^2\;\phi'}{a^4}\;\delta\sigma \, ;\\[.2cm]
&
  \delta \sigma = 2\,\plm^2 \, \frac{M_0  }{\phi '}\;\left[
  \Psi+\frac{\pi_0'}{\phi'}+c_b^2\;(3\; \Phi+k^2\;\pi_L)
  \right] \, ; \\ 
 &    -  v= \pi _L'+\frac{ \, M_1\;\phi' }{ 6\, a^2\,\mathcal{H}^2\,(1+w)}\;(\phi'\;\pi _L'-\pi_0)
   ,\qquad 
   \hat \Xi=-
   \frac{2  \; M_2  }{a^2} \;\pi_L \, . 
\end{split}
\ee
Then, comparing the perturbed Einstein equations \eq{eqEE}  with
 the expansion of the medium's perturbations in terms of the
 scalar fields \eq{eqPP},   we can read out directly the expression of the intrinsic entropic  perturbations  $\Gamma $ as a function of the entropy per particle fluctuation $\delta \sigma$
\ba
\Gamma\equiv\delta p- c_s^2\;\delta\rho\quad \Rightarrow\quad \Gamma=\frac{\phi '  \left( c_b^2  -  c_s^2\right)}{
   a^4}\;\delta\sigma \, .
\label{Gamma}
\ea
We stress that for barotropic fluids $ c_b^2  =  c_s^2 $ and
$\Gamma=0$.
The  evolution equation \eq{sig} for $\sigma$, expanded at linear
order, reads 
\be 
\delta \sigma'= \frac{M_1\plm^2}{  \phi'^2}\;\;k^2\;(\pi_0-\phi'\;\pi_L')
\label{sigk} \, .
\ee
Thus,  in the EFT approach for any medium, 
we have the following superhorizon limit $k/{\cal H}\to 0$
 \be
 \lim_{k\to 0}\delta\sigma=\delta\sigma_0(k)= \text{constant \,in\,time}  
\label{[sigmaconst}
\ee
Note that the above limit has to be taken carefully and, at the end, we have always to check that the equations for the background $\phi'$ and the perturbations for $\pi_{0,\,L}$ do not spoil such a statement
\footnote{Recently the violation of the Weinberg Theorem appears has
  been discussed in a number of papers, see for instance \cite{Akhshik:2015rwa,Akhshik:2015nfa,Chen:2013kta,Bordin:2017ozj}.
 The naive superhorizon limit $k\to 0$ actually means that the
 adimensional ratio $k/{\cal H}$ goes to zero. So,  for a given $k$
 mode, we have always  to check how  fast such a ratio  $k/{\cal
   H}$ decrease in time  with respect to other characteristic time
 scales of the problem.}.

Similarly,  for media such that $M_1=0$ we have $\delta
\sigma=\delta\sigma_0(k)=$ const. for any $k$. 
Referring to Table \ref{tab:femt}, this is the case for perfect fluids and solids.  On
the other hand, when $M_1\neq 0$ subhorizon entropic perturbations are
dynamically generated even if they were zero at superhorizon scales.

By using (\ref{me}) together with (\ref{eqPP}), (\ref{Gamma}) and
(\ref{sigk}), we arrive at the following coupled system of equations for
$\Phi$ and $\delta \sigma$
\ba\nonumber
\Phi'' &+& \left[ 3\,(1+c_s^2)\, {\cal H} + {\cal F}_1 \right] \Phi'
+\frac{M_2 \,\mathcal{H} \, \phi '}{6\, \plm^2\,a^4\, k^2 \,
   (w+1) \,\mathcal{H}^2}\, \delta \sigma'+\left[\frac{\phi' \,\left(c_b^2 -c_s^2\right)}{4\,  \plm^2 \,  a^2} + {\cal F}_2 \right] \delta \sigma +\\
&& \left[
  3\, \mathcal{H}^2 \,\left(c_s^2-w\right)+k^2\, c_s^2 + {\cal F}_3 \right]
\Phi =0 \,  ;
\label{eqPhi}
\ea
\ba \label{eqsigma}
&&\left [\frac{\phi'^2 \left[  a^2 \, \mathcal{H}^2 \, (w+1)+
   M_1\right]}{a^2   \, M_1 \,\plm^2\,\mathcal{H}^2\,   (w+1)}  \, \delta
\sigma' \right ]' +\left[
\frac{4 \, k^4  \,
M_2 \, \phi ' \, }{9\, \mathcal{H}^3\, (w+1)\,
   \left[\,a^2 \,k^2\, (w+1)+2 \,M_2\right]}\right] \, \Phi' -
 \\&&\left[\frac{2 \,k^4 \,  \,  \phi '
   \left(c_b^2-c_s^2\right)}{3\, \mathcal{H}^2 \, (w+1)} + {\cal F}_4\right] \Phi 
   -\left[\frac{\phi'{}^2 \,k^2\left[3 \, a^2 \, \mathcal{H}^2 \, (w+1)+M_0^2\,  ( c_s^2-2\,c_b^2  )\right]}{6 \,a^2 \,  M_0 \, \plm^2\,\mathcal{H}^2\,  (w+1)} + {\cal F}_5\,
\right] \delta \sigma \, =0 \, ;\nonumber
\ea
where
the ${\cal F}_i$ are functions of $k^2$, $M_i$ and such that 
\be\label{m20}
\lim_{M_2 \to 0} {\cal F}_{i} =0 \, , \qquad i=1,2,3,4,5 \, .
\ee
The explicit form of ${\cal F}_i$ are given in appendix \ref{sec:eqm}.
Equations \eq{eqPhi}-\eq{eqsigma} capture in a closed form the
dynamics of a general medium with 
$M_{1,0}\neq0$. When  $M_1\to 0$, $M_0\to 0$ Eq. \eq{eqsigma} is singular. 
The cases where $M_1$ and $M_0$ are zero deserve special scrutiny and
will be discussed in Section \ref{DoFsec}. 
The equation  for the Bardeen  potential \eq{eqPhi}  has  an
apparent singularity for $k\to 0$. 
However, Eq. \eq{sigk}  shows that  $\delta \sigma' \propto M_1\;k^2$,
as a result, the limit of $M_2\;\frac{\delta \sigma'}{k^2}\propto
M_1\;M_2$ for $k \to 0$ exists and  it is finite.

The source terms to the right-hand side of  \eq{me}  are important 
for the superhorizon 
evolution of the comoving curvature perturbation ${\cal R}$
defined as
\be
{\cal R}=
    -\Phi-\frac{2\;(\Phi'+{\cal H}\;\Psi)}{3\,{\cal
    H}(1+w)}.
\ee
Indeed, by using (\ref{me}), we get that 
\be
 6\;(1+w)\, {\cal H } \, {\cal R}' =  a^2 \, \hat{\Gamma} +
 k^2 \left(\frac{2\;\hat{\Xi}}{3} + 4 \, c_s^2 \, \Phi
 \right) \, .
\label{Req}
\ee
A similar analysis applies for the perturbation $\zeta$ of the curvature of constant
 energy density  hypersurfaces given by 
\be\label{zeta}
\zeta=-\Phi-\frac{\delta\rho}{3\;(\bar{\rho}+\bar{p})}={\cal
  R}-\frac{2\;k^2\;\Phi}{9\;(1+w)\;{\cal H}^2}=
-\Phi-\frac{a^2\;\delta\rho}{18\, \plm^2\;\mathcal{H}^2\;(1+w)}
\, .
\ee
We get for the  evolution of $\zeta$ 
\be\label{zeta1}
6\;(1+w)\;{\cal
  H}\;\zeta'=a^2\;\hat\Gamma+2\;k^2\;(1+w)\;\zeta+\frac{2\;k^2}{9\;{\cal
    H}^2}\; \left[2\;k^2+9\;(1+w)\;{\cal H}^2 \right]\;\Phi \, .
\ee
One  of the virtues of our effective field theory analysis is that it
provides $\Gamma$ and $\Xi$ as a function of the Bardeen potential and the entropic perturbations, namely   
\ba
&&\Gamma  =\frac{\phi '  \left( c_b^2  -
    c_s^2\right)}{a^4}\;\delta\sigma  \, ; \\
 && \hat \Xi=-\frac{2\;M_2}{a^2}\;\pi_L= M_2 \; 
 \frac{2\; a^2 \;\left[\left(2 \;k^2+3 \;(3 \;w+5) \; \mathcal{H}^2\right)\;\Phi  +6\; \mathcal{H} \;\Phi' \right]-\delta \sigma\; 
   \phi '}{3\; a^2\; \mathcal{H}^2 \;\left[a^2\; k^2 \;(w+1)+2\;
   M_2\right]} \, .
\label{anilim}
\ea
Thus, for  a general medium, there are two
sources which can trigger  a non-trivial dynamics for superhorizon perturbations
\cite{Weinberg:2003sw}: intrinsic entropic perturbations $\Gamma$ and anisotropic
stress  $\Xi$ when are non-vanishing for $k/{\cal H} \to 0$. \\
Anisotropic stress $\Xi$ is absent when $M_2 = 0$,
while intrinsic entropic pressure perturbations $\Gamma$ are ineffective,
see  Eq. \eq{dpp}, either for $c_b^2- c_s^2\propto \left.\frac{\partial \bar p}{\partial
    \bar\sigma}\right |_{\bar{\rho}}= 0$  or  $\delta \sigma =0$. For
barotropic fluids where $p=p(\rho)$, we have  $c_b^2-
c_s^2=0$. 
Generically, entropy perturbations are absent either when $M_0$ or  when $M_1=0$ with a vanishing initial value for $\delta \sigma$  (see Section \ref{sect:special}). 
For the anisotropic stress we note that,   taking the superhorizon limit,
one should specify the relative size of the different scales entering
in the game; namely we encounter in the dynamical equations
(\ref{eqPhi}-\ref{eqsigma}) contributions of the form
\be
\label{kto0}
k^2\;\Xi  \simeq
\begin{cases}
\frac{M_2}{{\cal H}^2\;(1+w)} \; 
 \left(\frac{2   \;(3 \;w+5) }{ a^2}\; \mathcal{H}^2 \;\Phi  + 4\;
  \frac{\mathcal{H}}{a^2} \;\Phi'  -\frac{\delta
  \sigma\;\phi'}{3\;a^4}\right) & M_2\ll a^2 \,k^2\;(1+w)\ll a^2\,{\cal H}^2\;(1+w)\\
 \frac{k^2}{{\cal H}^2}\;
 \left[   (3 \;w+5) \; \mathcal{H}^2 \;\Phi  +2\; \mathcal{H} \;\Phi'  -
 \frac{\delta \sigma\; 
   \phi '}{3\; a^2 } \right] &  a^2 \,k^2\;(1+w)\ll \{M_2,\;a^2\,{\cal H}^2\;(1+w)\}
\end{cases} \, .
\ee
This shows concretely how, in presence of different scales,  
sending to zero different dimensional   quantities ($k$ and $M_2$) do
not  commute: $[\lim_{k\to 0},\;\lim_{M_2\to 0}]\neq0$.
We will defer the detailed  study  of such  limits, related
to the   violation of  Weinberg theorem on the existence of adiabatic
modes~\cite{Weinberg:2008zzc,Weinberg:2003sw}, to a dedicated paper. 

\subsection{Tensor Perturbations}

Tensor perturbations are particularly simple, in fact the   transverse and traceless spin two part
$\chi_{ij}$ of the metric perturbations
\be
ds^2 = a^2 \left[ -dt^2 +\left(\delta_{ij}  + \chi_{ij}(t,\,\vec x )\,\right) \;\; dx^i
  dx^j \right] \, .
\ee
are  gauge invariant. The quadratic Lagrangian for  tensor perturbations  in the
Fourier basis is \cite{Dubovsky:2004ud,Rubakov:2008nh,Blas:2009my,ussgf}
\begin{equation}
L^{(2)}_t=\frac{\plm^2}{2} \left[a^2
\; \chi_{ij}'^2- \chi_{ij}^2 \;\left(k^2 \;a^2+M_2\right)\right ] \, .
\end{equation}
Thus, the linearised Einstein equations for the tensor modes reads
\be
\chi_{ij}'' + 2 \, {\cal H} \, \chi_{ij}' + \left( k^2 + \frac{M_2}{a^2} \right)
\chi_{ij} =0 \, .
\ee
For fluids and superfluids where $M_2=0$, the dynamics of spin 2
modes is standard. This is not the case for solids and supersolids
where $M_2 \neq 0$ and it can trigger a enhancement/suppression of
$\chi_{ij}$ depending on its sign. 
This could induce, for instance, observable effects on the propagation and lensing of CMB B-modes if the continuous medium is relevant at sufficiently early times.
Remarkably the mass parameter, $M_2$,  responsible for the
gravitational slip $\Phi - \Psi$,  also enters in the propagation of gravitational waves.

\subsection{Vector Perturbations}
In the vector sector is more convenient to use the
unitary gauge and set $\pi^i_V=0$, while the  perturbations of the metric have the form  
\be
ds^2 = a^2 \left[ -dt^2 +\left(\delta_{ij}  + \partial_i\,s_j(t,\,\vec x)+  \partial_j\,s_i(t,\,\vec x)\right) \, dx^i
  dx^j +2\;\n_i(t,\,\vec x) \;dt\;dx^i\right]  \, ;
\ee
 with $ \partial_i \,s_i=\partial_i \,\n_i=0$.
 The quadratic Lagrangian reads \cite{Dubovsky:2004ud,Rubakov:2008nh,Blas:2009my,ussgf}
\ba
L^{(2)}_v&=\frac{\plm^2}{2}  \left[k^2 a^2 \left(\n_i-\;s_i'\right)^2-k^2 M_2\;s_i^2+M_1^{eff} \n_i^2\right]
\, ;
\ea
where
\ba\label{m1eff}
M_1^{eff} \!\!&\equiv&\!\!  M_1+\frac{a^4  (\bar{\rho}+\bar{p})}{\plm^2}=
M_1+6\, a^2 \, \mathcal{H}^2(1+w)
=\frac{\lambda_1^2}{\phi'}+ \frac{a^4\;\bar{\rho} }{\plm^2} \nb \\
\!\!& =&\!\! \frac{1	}{\plm^2} \left(2\, \phi'^2\,  \sum_{n=0}^3\;a^{-2n}\;U_{y_n}+ \phi'\;
a^3\;U_Y-a\;U_b-2\;\sum_{n=1}^3\;n\;a^{4-2\,n}\;U_{\tau_n} \right)\, .
\ea
The fields $\n_i$ have a purely algebraic  equations of motion
\ba
\n_i=\frac{k^2 \;a^2\; s_i'}{k^2 \;a^2+M_1^{eff} } \, ,
\ea
 and thus they can be integrated out, giving the Lagrangian
\ba
L^{(2)}_v&=\frac{\plm^2}{2}\; k^2 \left(\frac{ a^2\;M_1^{eff} }{ k^2 \;a^2+ M_1^{eff}}\;s_i'^{\; 2}- M_2\;s_i^2 \right)\, .
\ea\
The vector $s_i$ propagate only if $M_1^{eff}\neq 0$.
The dispersion relation is not trivial only when $M_2\neq  0$. 
Thus, $M_2$, besides controlling the dispersion relation of tensors, also determines the dynamics of vectors. 

\section{Masses and Degrees of Freedom}
\label{DoFsec}
In order to disentangle the two  $M_{1,0}= 0$ it is
convenient to examine the structure of the equations of motion retaining all
the original fields, though some of them can be integrate out.
From  $\delta \rho$ in \eq{eqEE}, \eq{eqPP} and
  \eq{anilim}, we get the relation
\ba\label{pol1}
&&\pi_L= M_2\;\left(\# \;\delta\sigma+\# \;\Phi+\# \;\Phi' \right)\,  \ea
From  \eq{sigk}   and the derivative of \eq{pol2} we obtain
\ba\label{pol2}
&& \delta\sigma'=M_1\;\left( \#\;\pi_0+ \# \;M_2\;\delta\sigma+\#
  \;\Phi+\# \;\Phi' \right)  \, ;  \ea
 Finally, the definition of $\delta \sigma$ in \eq{eqPP}, together with\eq{pol2},  give
  \ba \label{pol3}
&& \delta\sigma=\left[\# \;M_0\;\pi_0'+(\#\;M_4+\#\;M_0)\;( \Phi+\#
  \;\Phi')\right] \, .
\ea
We denote
with  $\#$  a generic  functions of $k,\;{\cal H},\; a, \; M_i$
whose detailed form is not relevant for us.
The above  set of equations is equivalent to the coupled  system of second order
 differential  equations for $\Phi$ and $\delta \sigma$ given
in (\ref{eqPhi}-\ref{eqsigma}) and  in appendix \ref{sec:eqm}.
Let us now examine the following degenerate cases.
\begin{table}[h!!]
  \footnotesize
  \small
  \renewcommand\arraystretch{2.2}
  \centering
  \begin{tabular}{|c|c|c|c| c|}
    \hline
    $M_0$ &  $M_1$&  $M_1^{eff}$&  Propagating\;DoF    & Eqs  for $\;\Phi$ and $\delta\sigma$\\
    \hline \hline
    $\neq0$      &       $\neq0$         &      $\neq0$              &
                                                                       $\Phi,\;\pi_0$
                                                                       (or $\pi_L$)
                                                                       
                                                       & $\Phi''+...=0,\;\delta \sigma''+...=0$\\ \hline
        0     &            $\neq0$    &          $\neq0$          &     $\Phi$              & $\Phi''+...=0,\;\delta \sigma =0$\\ \hline
          $\neq0$      &         0     &        $\neq0$            &
                                                                     $\Phi,\;\pi_0$  
                                                       &$\Phi''+...=0,\;\delta \sigma'=0$ \\ \hline
         $\neq0$       &   $\neq0$             &            0      &      $\Phi$            &$\Phi''+...=0,\;\delta \sigma+...=0$\\ \hline
                        \end{tabular}
  \caption{\it \small Structure of the scalar equations of motion and degrees
    of freedom (DoF) in terms of the masses \label{tab:prop}}
 \end{table}
\begin{itemize}
\item $M_1$=0.  

\no
From \eq{pol2} we have that 
\be\label{ad}
\delta\sigma'=0 \;\;\to\;\;\delta \sigma=\delta\sigma_0(k).
\ee
Then \eq{pol3}  becomes a  second  order  equation  for  $\pi_0$.
On the contrary, \eq{pol1} shows that $\pi_L$ is an auxiliary field.
Thus, there are two propagating fields:  $\pi_0$ and $\Phi$.
\item $M_0$=0 

\no
Notice that \eq{backg} implies that also $M_4=0$. From \eq{pol3} it
follows that 
\be\label{iso}
\delta \sigma=0 \, .
\ee
As a result,  \eq{pol1}-\eq{pol2} imply that both $\pi_0$ and $\pi_L$
are auxiliary fields.  Thus the only propagating field is $\Phi$ \cite{Comelli:2013tja,Comelli:2013txa,Comelli:2013paa,Comelli:2012vz}.
\end{itemize}
Thus, irrespective of the values of $M_{0,1}$,  $\pi_L$ is  an
auxiliary field \eq{pol1} and can be  always integrated out. Moreover,   $\pi_0$ can be
traded for the gauge invariant entropy per particle perturbation $\delta \sigma$.
Let us consider the dynamics of $\delta \sigma$. The coefficient of
$\delta \sigma'$ in \eq{eqsigma} is proportional to $M_1^{eff} $.
 The following case is possible.
\begin{itemize}
\item    $M_1^{eff}=0$ is special: $\delta \sigma$  has to satisfy \eq{eqsigma}
\be
\#\;\delta \sigma+\#\;\Phi'+\#\;\Phi=0 \, ;
\ee
i.e. $\delta \sigma$   is determined by $\Phi$
and from \eq{pol1}, \eq{pol2} we see that both $\pi_L$ and $\pi_0$ are
auxiliary fields. Thus,  again only $\Phi$ propagates. In such a case also vectors do not propagate and we have a total of three degrees of freedom (one scalar and two tensors) \cite{Comelli:2014xga}.
Note the difference with the case $M_1=0$, where though still $\delta
\sigma'=0$ and there is an extra  propagating scalar mode.\\
\end{itemize}
A summary of the above results is given in Table \ref{tab:prop}.\\
Let us briefly discuss the connections between massive gravity
theories and self gravitating media. In~\cite{ussgf} it was shown
that rotational invariant massive gravity theories, described by the potential~\cite{Dubovsky:2004sg,Rubakov:2008nh,Comelli:2014xga,Lin:2015cqa} $V(g^{00},\;g^{0i},\;g^{ij})$,
are equivalent, up to a gauge transformation,  to a medium described by the Lagrangian  $U(b,\,Y,\,X,\,\tau_n,\,y_n)$. 
\\ 
In massive gravity the existence of a scalar sixth mode is typically
associated to a ghost mode which  generates instabilities at any scale.
It is worth to stress that   even   media like perfect
fluids have  six degrees of freedom, supporting a second scalar mode without any apparent instability~\cite{Comelli:2014xga}.
Let us consider for instance the perfect fluid  $U(b,\,Y)$.
Once the variables $\Phi$, $\delta \sigma$ are used,  the equations of
motions \eq{mee} take a very simple   form  (we set $c_s^2=w$ for simplicity) 
\be
\Phi''=-k^2\;w\;\Phi-3\;{\cal
  H}\;(1+w)\;\Phi'+\frac{w-c_b^2}{4\;a^2\;\plm^2}\;\phi'\;\delta\sigma_0
\, 
\ee
and the conservation of entropy per particle $\delta\sigma'=0$ implies $\delta \sigma=\delta \sigma_0$. No
sign of instabilities is present, as it should  be, being a
perfect fluid.  
The matter will be studied in a future dedicated paper.

 \section{Adiabatic and Isentropic Media}
\label{sect:special}
 
 Adiabatic media feature a constant in time entropy per particle
 $\sigma(\vec x)$. From \eq{ad}, we see that this happens whenever
 $M_1=0$. Thus, at least at the linearised order, adiabaticity is
 equivalent to 
\ba\label{m1}
M_1\equiv \frac{2\, \phi'^2}{\plm^2} \left(\sum_{n=0}^3 \;  a^{-2 \;n}\;
   U_{y_n}+a^2\;  U_{X}\right) =0 \, .
\ea
Clearly, a sufficient condition for adiabaticity is the absence in the
effective action (\ref{LAll}) of $X$ and $ y_n$. As will see later
this is the case for perfect fluids and solids.

A stronger thermodynamical requirement is that the medium is isentropic, namely
$\sigma$ is  strictly a constant and thus no temporal or spatial
variations are allowed. This implies that $\delta \sigma$ 
should vanish identically. From  \eq{iso} this is the case when 
\ba
\label{m0}
M_0\equiv \frac{ \phi'^2\;a^2}{2\plm^2}\;
   \left[ \left(U_{Y^2}-2\;
   U_X\right)-4\;\frac{ \phi '}{a}\; U_{YX}+4 \;\frac{\phi'{}^2}{a^2}\;U_{X^2} \right] =0  \, .
\ea 
A sufficient condition for an isentropic medium is that  the function $U$ entering
the effective action (\ref{LAll}) does not depend from the operators
$Y$ and $X$. From our thermodynamical dictionary, see (\ref{ch1}), it
is clear that  the presence of the operator $Y$
turns on the entropy density $s$, while $X$ can be related to a superfluid component.
 
From the background equations of motion \eq{backg}, it follows that
requiring   $M_0=0$ implies, for consistency,   also that $M_4=0$;  in
this case, the symmetries of the medium are enhanced,  
the $\phi$ becomes a gauge artefact and one can set $\phi'=1$.
Moreover, both $c_s^2$ and $c_b^2$ in  Eq. \eq{backg} are singular; actually we have that
\be\label{css}
c_s^2=\frac{M_2-3 \;M_3}{9\; a^2 \, \mathcal{H}^2(1+w)} \, .
\ee
From \eq{pol3} we have  $\delta \sigma=0$  and the fluid is isentropic.

A medium can be isentropic even when $U$ depends on $Y$ and $X$  for
a suitable choice of $U$. Interpreting  \eq{m0} as a differential
equation for $U$ in the $X$ and $Y$ variable, one can verify that for instance the Lagrangian
$\sqrt{-X}\;{\cal U}_1(b,\,\tau_n,\,y_n)+Y\;{\cal
  U}_2(b,\,\tau_n,\,y_n)$ is isentropic. This is the case also for 
$U(\frac{X}{Y^2},\,b,\,\tau_n,\,y_n) $  where $Y $ and $X$ appears in
the special combination $\frac{X}{Y^2}$ typical of the operators
${\cal O}_\a$ entering in a  subclass of superfluids and
supersolids with $U({\cal O}_{\a\b n},\,\tau_n)$, see \eq{myn}.
Finally, also $U(X+Y^2,\,b,\,\tau_n,\,y_n) $  forms a rather general class of  isentropic media.
The combination $X+Y^2$ is precisely the thermodynamical variable entering in the description of superfluids and supersolids.

\section{Perfect Fluids}
\label{sect:fluids}
Perfect fluids are probably the simplest media one can think of and
are ubiquitous in cosmology. They are 
characterised by an EMT  (see \eq{sgmemt}) with vanishing  heat flow $q_\mu$ and  anisotropic stress
    ${\cal P}_{\mu\nu}$ \cite{Andersson:2006nr,Weinberg:2008zzc}. Thus the pressure is isotropic and $M_2=0$, being   $\Xi=0$, see \eq{eqPP}. The internal symmetry for  perfect fluids  corresponds to
spatial volume preserving diffeomorphisms and, as shown in Table
\ref{tab:summa}, this select at  leading order a small number of
operators: $b$, $Y$ and $X$.
From the symmetry requirements and by inspection of the general form
of the EMT \eq{eqO}, one can deduce that $U(b,\;Y)$  describes
a perfect fluid \cite{Endlich:2010hf,Dubovsky:2011sj,Ballesteros:2012kv,Nicolis:2013lma,Ballesteros:2014sxa,Delacretaz:2014jka}. Having $M_1=0$, the entropy per
particle is conserved.  
The  equations of motion for $\Phi$ and $\delta \sigma$ read
\ba
\label{mee}
 &&\Phi '' = \frac{ 
   \left(c_s{}^2-c_b{}^2\right) \phi
   '}{4\; a^2 \;\plm^2}\delta \sigma_0+  \left[3\,
   \mathcal{H}^2\;
   \left(w-c_s{}^2\right)-k^2\;
   c_s{}^2\right]\;\Phi-3\;
   \left(c_s{}^2+1\right)\; \mathcal{H}\;
   \Phi ' \, ; \\[.2cm] 
&& \delta\sigma =\delta\sigma_0(k) \, .
\ea
Note that for barotropic fluids described by $U(b)$ and $U(Y)$ we have
$c_b^2=c_s^2$; thus  no entropic source  is present in the evolution
equation  for the Bardeen  potential.
A perfect fluid can be also  described by  $U(X)$ \cite{Matarrese:1984zw}; however, in this
case  $M_1^{eff}$ =0, and, as shown in Section \ref{DoFsec}, the field
$\delta \sigma$ is non dynamical.  Moreover, from \eq{eqsigma}, taking
the  limits: $M_1^{eff}\to 0,\;M_2\to 0$ and $c_b^2=c_s^2$, we get
that $ \delta \sigma  \, =0$, so the fluid is also isentropic. \\
The main features of perfect fluids are summarised  in Table \ref{tab:PF}.
\vskip .5cm
\begin{table}[h!]
  \footnotesize
  \small
  \renewcommand\arraystretch{2.2}
  \centering
  \begin{tabular}{|c|c|c|c|c|c|c|c|c|}
    \hline
Lagrangian            &  $M_0$ &  $M_1$&    $M_2$&  $M_3$&  $M_4$&  $M_1^{eff}$ & DoF & Features\\ \hline \hline
$U(b)$                &  0         & 0        &    0          &   $\neq 0$        &          0   &  $\neq 0$ & 1& Barotropic, Isentropic \\ \hline
$U(Y)$                &   $\neq 0$         &      0             &0          &  0         &      $\neq 0$          &  $\neq 0$ & 2 & Barotropic, Adiabatic \\ \hline
$U(b,\;Y)$            &       $\neq 0$       & 0  &0          &  $\neq 0$           &     $\neq 0$             &   $\neq 0$ &2 & Adiabatic \\ \hline
$U(X)$                &       $\neq 0$       &     $\neq 0$  &0         & 0          &   0         & 0  & 1& Barotropic, Isent.,  Irrot.\\ \hline 
\end{tabular}
  \caption{\it \small Masses and thermodynamical classification  of Perfect Fluids \label{tab:PF}}
 \end{table}
 
\no

\section{Superfluids}
\label{sect:superfluids}

Superfluids are characterised by the EMT in \eq{sgmemt} with $q_\mu=2\,Y\,U_X\,\xi_\mu$ and ${\cal P}^{tl}_{\mu\nu}=-2\;U_X\;\xi_\mu\;\xi_\nu
$. 
Moreover,  the anisotropic perturbation $\Pi_{\mu\nu}$ assumes the specific form $\Pi_{\mu\nu}=-2\,U_X\, \xi_\mu\,\xi_\nu $.
A superfluid can be
roughly thought has a mixture of a perfect fluid plus a superfluid
irrotational component and can be described  by the Lagrangians
$U(b,\;Y,\;X)$ or $U(b,\;{\cal O}_\a)$ \cite{Son:2002zn,Nicolis:2011cs,Nicolis:2013lma,Delacretaz:2014jka}. 
Superfluids, besides spatial volume preserving diffs,
support a 
 temporal shift symmetry $\varphi^0\rightarrow\varphi^0+c$, where $c$ is constant. 
Around a  FLRW background, being the relative velocity $\xi_\mu$ of order one, it implies that ${\cal P}^{tl}_{\mu\nu}\propto \xi_\mu\;\xi_\nu $ is at least second  order in cosmological perturbation theory.
 Indeed, we always have $M_2=0$. 
Then, the equation for the Bardeen potential $\Phi$ is (see the limits \eq{m20})
\ba\label{sup}
 \Phi '' &=& \frac{
   \left(c_s{}^2-c_b{}^2\right)\; \phi
   '}{4\; a^2\; \plm^2}\;\delta \sigma+\left[3\;
   \mathcal{H}^2\;
   \left(w-c_s{}^2\right)-k^2\;
   c_s{}^2\right]\;\Phi  -3
   \left(c_s{}^2+1\right)\; \mathcal{H}\;
   \Phi '  \, .
   \ea
The entropy per particle   is not conserved ($M_1\neq 0$) and satisfies
the following equation 
\be
   \left[\frac{\phi'^2 \, 
   M^{eff}_1 }
   {6 \, a^2 \,  M_1 (w+1) \, 
   \mathcal{H}^2} \delta \sigma'\right]'=
 k^2 \,  \left[
   \frac{\phi'^2 \, 
   (c_s^2-2\,
  c_b^2)}{6 \, a^2 \, (w+1)\,
   \mathcal{H}^2}+
   \frac{ \phi'^2}{2 \, M_0}
   \right]\delta\sigma
  + 
   \left[
   \frac{2  \,\plm^2\;  k^4\,  \phi'
   \,(c_b^2-c_s^2)
   }{3 \, (1+w)\,  \mathcal{H}^2}
   \right] \,  \Phi.
\ee
The main features of Superfluids are summarised in Table \ref{tab:SS}.
\begin{table}[ht]
  \footnotesize
  \small
  \renewcommand\arraystretch{2.2}
  \centering
  \begin{tabular}{|c|c|c|c|c|c|c|c|c|}
    \hline
Lagrangian            &  $M_0$        &  $M_1$      &   $M_2$   &  $M_3$        &  $M_4$&  $M_1^{eff}$ & DoF &Feature\\ \hline \hline
$U(b,\;X)$         &  $\neq 0$         &  $\neq 0$   &    0      &     $\neq0$      &  $  \neq0$   &     $\neq0 $ &  2 & \\ \hline
$U(Y,\;X)$         &   $\neq 0$        &  $\neq 0$   &   0        &     0              &  $ \neq0$   &     $\neq0 $ &  2 & \\ \hline
$U(b,\;Y,\;X)$      &   $\neq 0$       &  $\neq 0$   &    0      &     $\neq0$   &  $  \neq0$   &     $\neq0 $ &  2 &\\ \hline
$U({\cal O}_\a)$      &   $  0$       &  $\neq 0$   &    0      &     $ 0$   &  $   0$   &     $\neq0 $ & 1 & Isentropic\\ \hline
$U(b,\;{\cal O}_\a)$      &   $  0$       &  $\neq 0$   &    0      &     $\neq 0$   &  $   0$   &     $\neq0 $ &  1 & Isentropic\\ \hline
\end{tabular}
  \caption{\it \small Masses and features of superfluids \label{tab:SS}}
 \end{table}
 
\no
Isentropic superfluids with Lagrangian $U(b,\,{\cal O}_\a)$ are  rather
peculiar. The form of $U$ is protected by symmetry, see
Table \ref{tab:summa}, and only a single scalar degree of freedom is present. Indeed,
in the combination $X/Y^2$ the $\pi_0$ field doesn't enter at the first order, see 
 \eq{byx}. Finally Eq. \eq{mee} with $\delta\sigma_{0}=0$ shows that  such
 class of  media behaves more like isentropic
 perfect fluids rather then  superfluids.
Superfluids in cosmology  are typically associated to    the Lagrangian $U(X)$ 
   with possible shift symmetry breaking
\cite{Berezhiani:2015bqa,Cai:2015rns} with possible   connections  with
Mond \cite{Milgrom:1983ca};  for a
recent analysis, also  with other operators, see \cite{Berezhiani:2015pia}.     

\section{Solids}
\label{sect:solids}

Solids are described by the Lagrangian $U(\tau_n)$ \cite{Carter:1987qr,Kijowski:1997mx} or, for  finite
temperature solids, by $U(Y, \tau_n)$. 
Differently from fluids,  the anisotropic stress $\Pi_{\m\n}$ is not
vanishing which implies that the two Bardeen potentials, $\Psi$ and
$\Phi$, are not equal \eq{eqEE}
\be
 \Phi - \Psi = \frac{\Xi}{2\;\plm^2}=-\frac{M_2}{\;a^2}\;\pi_L \, .
\ee
 Being $M_1=0$, solids are adiabatic, thus $\delta \sigma=\delta
 \sigma_0(k)$. 
 The evolution equation for $\Phi$  reads
\be
\begin{split}
\Phi'' +& \left[ 3\;(1+c_s^2)\; {\cal H} + {\cal F}_1 \right] \Phi'
   +\left[\frac{\phi' \left(c_b^2 -c_s^2\right)}{4\,  \plm^2 \,  a^2} + {\cal F}_2 \right] \delta \sigma_0  + \left[
 3\, \mathcal{H}^2 \left(c_s^2-w\right)+k^2 c_s^2 + {\cal F}_3 \right]
\Phi =0 \, .
\end{split}
\label{eqsol}
\ee
Notice that for  solids described by $U(\tau_n)$ we also have  $M_{0,4}=0$ so that $\delta \sigma=0$,
$c_b^2 $ is not defined while for the speed of sound $c_s^2$ we have that
\eq{css} is still valid.

\begin{table}[ht]
  \footnotesize
  \small
  \renewcommand\arraystretch{2.2}
  \centering
  \begin{tabular}{|c|c|c|c|c|c|c|c|c|}
    \hline
Lagrangian            &  $M_0$ &  $M_1$&   $M_2$  &  $M_3$        &  $M_4$&  $M_1^{eff}$ & DoF& Features\\ \hline \hline
$U(\tau_n)$         &  0         & 0        &    $\neq 0$  &     $\neq0$   &  $  0$   &     $\neq0 $ &  1& Isentropic \\ \hline
$U(\tau_n,\;Y)$    & $\neq 0$ & 0     &  $\neq 0$    &      $\neq 0$  & $\neq 0$ & $\neq0 $ &2 & Adiabatic  \\ \hline
\end{tabular}
  \caption{\it \small Mass spectrum and thermodynamical classification of Solids \label{tab:S}}
 \end{table}
The properties of solids are summarised in Table \ref{tab:S}.
Superhorizon perturbations for solids are similar to  supersolids due to the fact that in such a limit
  all fluids allow adiabatic solutions.
Solids received lot of attentions in many cosmological contests, see  
\cite{Bucher:1998mh,Battye:2005hw,Battye:2013er,Pearson:2014iaa,Balek:2014uua,Kang:2015uha,Alberte:2015isw,Bordin:2017ozj}.

 \section{Supersolids}
\label{sect:supersolids}

 Supersolids are characterised by the presence of unrelated 
relative velocity $\xi_\mu $ and anisotropic perturbation tensor $\Pi_{\m\n}$ \cite{Son:2005ak, Nicolis:2013lma,Delacretaz:2014jka}.
General supersolid have two scalar degrees of freedom described by the coupled set of equations (\ref{eqPhi}) and (\ref{eqsigma}).
Let us discuss briefly the superhorizon  ($k\to0$) regime.  
The limit for  small $k/{\cal H}$ can be implemented in  two different ways as show in \eq{kto0}
depending on the relative size between 
$k^2$  and $M_2$. For simplicity we set here $c_s^2=w$.  
For $M_2\ll
a^2\,k^2\,(1+w)\ll  a^2\;{\cal H}^2\,(1+w)$  we get %
\ba
\Phi''=-3\; (w+1)\;
   \mathcal{H}\; \Phi '+
\frac{\delta \sigma \;\phi '\;
   (w-c_b^2)}{4 \,a^2\, \plm^2} \, 
   \label{ph2s}
\ea
and in absence of entropy perturbations, the Bardeen potential   $\Phi$ at leading order is constant.
\\
In the second case,  $a^2\,k^2\,(1+w)\ll \{a^2\;{\cal H}^2\,(1+w),M_2\}$, setting
$M_2' = \;{\k}_2 \;{\cal H}\;M_2$, we have  
\ba\nonumber
\Phi'' &=&-\delta \sigma \; \left(\frac{\phi ' \left(3\;(
  c_b^2-w)+\kappa _2 -2\right)}{12\;
   a^2\; \plm^2}-\frac{M_2\;\phi'}{18\;M_{Pl}^2\;(1+w)\;a^4\;{\cal H}^2}\right)+ \mathcal{H} \;\left(\kappa _2-3\;
   w-5\right)\; \Phi ' +\\&&\label{ph1s}
   \left(
   \frac{1}{2} \left(\kappa
   _2-2\right) (3 \;w+5) \;\mathcal{H}^2 -\frac{M_2}{a^2}
    \right)\;\Phi
   \ea
This time the Bardeen  potential is not constant also in absence of
entropy perturbations. 
The above observation has interesting implications for inflationary
models where superhorizon perturbations violate the adiabatic Weinberg theorem \cite{Endlich:2012pz, Kang:2015uha, Bartolo:2013msa,
 Akhshik:2014gja,Bartolo:2014xfa,Ricciardone:2016lym}.
 In a dedicate paper we will analyse the behaviour of the various gauge invariant scalars in the superhorizon limit.
\\ 
There are also some special supersolids that deserve a mention as the subclasses 
  $U(X,\;w_n)$, $U(w_n)$ and $U({\cal O}_{\a\b n})$ that result  symmetry
  protected, see Table \ref{tab:summa}.
  The first two Lagrangians  have   $M_1^{eff}=0$ and, from the analysis of Section \ref{DoFsec}, it follows
  that $\delta \sigma$ is an auxiliary field and only a single scalar
  degree of freedom is present.
 The supersolids $U({\cal O}_{\a\b n})$  have $M_0=0$ \eq{myn} and so they are isentropic.

\section{Conclusions}
\label{sect:conc}
We   have studied the dynamics of cosmological perturbations around a
FLRW Universe in the presence of a generic self-gravitating medium
  by using an effective field theory approach. The low
energy modes  of the medium (phonons) are related to four scalar
fields $\varphi^A$ corresponding to the Goldstone modes of the  spontaneously
broken  spacetime translations.  
We provide a complete classification of the medium both from a dynamical than from a  dynamical and thermodynamical  point of view.
The Lagrangian density $U$ describing the medium depends on a set of   scalar
operators built from the scalar fields $\varphi^A$ according to  prescribed
internal symmetries.  
The dynamics of scalar perturbations is generically described  by  two
coupled  second order 
differential 
equations, one  for the Bardeen potential $\Phi$ and one for the fluctuations
of the entropy per particle $\delta \sigma$, that  turns out to be a
combination of Goldstone fields.
 Besides 
the background pressure and energy density, the dynamics of linear cosmological
perturbations is encoded in a  set of five   masses $\{M_i \}$ derived from the medium's Lagrangian.
Media are classified  according to the internal symmetries of the EFT
and the structure of the  EMT which impact on the values of  $\{M_i \}$.
Remarkably, we find that also simple thermodynamical properties of
a medium correspond to specific values of the mass parameters $\{M_i \}$.
We have found that media can be classified according to the following scheme:
\begin{itemize}
\item Adiabatic media, with $\delta\sigma(\vec x)$ time
  independent, have $M_1=0$:
\begin{itemize}
\item Perfect Fluids at finite Temperature: $U(b,\,Y)$
\item Solids at finite Temperature: $U(\tau_n,\,Y)$
\end{itemize}
\item Isentropic media with $\delta\sigma=0$  are characterised
  by $M_{0}=0$:%
\begin{itemize}
\item Perfect Fluid: $U(b)$ 
\item Solids: $U(\tau_n)$  
\item Superfluid: $U({\cal O}_\a),\;U(X+Y^2)$
\item Supersolids:  $U({\cal O}_\a,\,\tau_n,\;y_n),\;U({\cal O}_{\a\b n}),\;U(X+Y^2,\,\tau_n,\;y_n),\;\sqrt{-X}\;{\cal U}_1(\,\tau_n,\;y_n)+Y\;{\cal U}_2(\,\tau_n,\;y_n)$
\end{itemize}
\item The  Lagrangian $U(X)$ describes an   irrotational isentropic
   perfect fluid with $\delta\sigma=0$; indeed
    $M_1^{eff}=M_2=0$ and 
$c_b^2=c_s^2$.
\item Isotropic media have zero anisotropic stress $\Pi_{\m\n}=0$ and thus the two Bardeen potentials $\Phi$ and $\Psi$  are   equal. Such a media are characterised by $M_{2}=0$ and are: 
\begin{itemize}
\item Perfect Fluids;  
\item Superfluids.
\end{itemize}
\item 
Generically, superhorizon perturbations for all media admit
adiabatic solutions ($\lim_{k/{\cal H}\to 0}\delta\sigma(k,t)=\delta\sigma(k)$) despite,  in superfluids and supersolids, entropy  
perturbations have a non trivial dynamics. 

\item Media which are non adiabatic but with still non-dynamical entropy perturbations are characterised by   $M_1^{eff}=0$ with Lagrangian 
$U(b\;Y,X,\;w_n)$. For such media the Bardeen potential $\Phi$ determines
completely $\delta\sigma$, namely $\delta\sigma=f(\Phi)$.

\end{itemize}

As shown by \eq{ph2s}-\eq{ph1s}, the superhorizon evolution of the
Bardeen potential can be 
not trivial.  Recently in a number of models
\cite{Namjoo:2012aa,Chen:2013aj,
  Chen:2013kta,Akhshik:2015nfa,Gruzinov:2004ty,Endlich:2012pz} it has been
reported  violations of   the adiabatic Weinberg
theorem~\cite{Weinberg:2003sw,Weinberg:2008zzc}  whereby curvature perturbations are constant on  super-horizon scales \eq{Req}.
The  violation of the theorem, which allows superhorizon modes to grow,
involves the analyticity properties of  the suitably normalized
Goldstone fields $\pi_{0,L}$ in the limit $k\to 0$ and the  existence
of peculiar backgrounds (see for instance  fluid inflation~\cite{Akhshik:2015rwa,Chen:2013kta}) for which
the would be decreasing mode is turned into a growing one.   In  our approach, we trade the Goldstone fields for the Bardeen
 potentials  and the entropy perturbation $\delta \sigma$. In such a
 way, the sources of the  
 violation of the adiabatic Weinberg theorem become evident from the
 equations \eq{ph2s} and \eq{ph1s}.
A complete analysis of such a matter, including the inflationary
backgrounds, for generic media deserves a dedicated study and it will
be given elsewhere.

\section*{Acknowledgements}
D.C.\ and L.P.\ would also like to thank G.\ Ballesteros for
enlightening discussions during the early stages of the project.

\begin{appendix}

\section{EMT and Currents}
\label{emta}
The EMT derived from the action (\ref{LAll}) is~\cite{ussgf}
\be
 T_{\mu\nu}=U\,g_{\mu\nu} -2\; \frac{\de U}{\de C^{AB}}\;\de_\mu \varphi^A \; \de_\nu \varphi^B=
U\, g_{\mu\nu}-
2\,\sum_k U_{\mathcal{O}_k}\, \frac{\partial \mathcal{O}_k}{\partial
  g^{\mu\nu}}\,,
\label{emtgen1}
 \ee
where $\mathcal{O}_k$ are the nine scalar LO operators appearing in
\eq{LAll} and we use the notation $U_{\mathcal{O}_k}=\partial
U/\partial \mathcal{O}_k$.  Their partial derivatives with respect to
$g^{\mu \nu}$ are 
\begin{equation}
\begin{aligned}
 \label{eqO}
\frac{\partial Y}{\partial g^{\mu\nu}}=-\frac{Y}{2}\;u_\mu\,u_\nu\,,\quad
  \frac{\partial X}{\partial g^{\mu\nu}}= -X\,\vel_\mu \vel_\nu\,,\quad
   \frac{\partial \tau_n}{\partial g^{\mu\nu}}=n\,\,\partial_\mu
   \boldsymbol{\varphi}
   \cdot\boldsymbol{B}^{n-1}\cdot\partial_\nu \boldsymbol{\varphi} \,,\quad 
\frac{\partial b}{\partial
  g^{\mu\nu}}=\frac{b}{2}\;h_{\mu\nu} \,, \\
   \frac{\partial y_n}{\partial g^{\mu\nu}}=\sum_{m=1}^{n}\partial_\mu
   \boldsymbol{\varphi} \cdot\boldsymbol{B}^{n-m}\cdot
   \boldsymbol{Z}\cdot \boldsymbol{B}^{m-1}\cdot \partial_\nu \boldsymbol{\varphi}
   -2\sqrt{-X} \, \boldsymbol{\mathcal{C}} \cdot
     \boldsymbol{B}^n\cdot \partial_{(\mu} \boldsymbol{\varphi} \,
     \vel_{\nu )}\,,   \quad\quad\quad\quad\quad
\end{aligned} \, .
\end{equation}
 where 
\be
{\cal C}^i \equiv C^{0i} \quad ,
\ee
and  $\boldsymbol{{\cal C}}$ denotes the vector with
components ${\cal C}^i $, in these expressions
 $\partial_\mu \boldsymbol{\varphi}$ has to be understood as a $1\times 3$ matrix of
 components $\partial_\mu\varphi^i$. We have used that 
\be
\frac{\partial u^\alpha}{\partial g^{\mu\nu}}=-\frac{u^\alpha}{2}\;u_\mu\,u_\nu\,,\quad \frac{\partial\mathcal \vel^\mu}{\partial g^{\alpha\beta}}=-\frac{\vel^\mu}{2}\vel_\alpha\vel_\beta\,.
\ee
The dot ($\cdot$) represents the
 standard three-dimensional matrix product. Notice that
for convenience we have included $b$, though it can  be written as a combination of
 the three $\tau_n$. 
The operators $w_n$ are not independent; they are non-linear combinations of the scalars $X$, $\tau_n$ and $y_n$:
\be
\label{Wn}
w_1=\,\tau_1-\frac{y_0}{X},\quad
w_2=\,\tau_2-2\,\frac{y_1}{X}+\frac{y_0^2}{X^2},\quad
w_3=\,\tau_3-3\,\frac{y_2}{X}+3\,\frac{y_0\,y_1}{X^2}-\frac{y_0^3}{X^3}
\, .
\ee 
The four Noether currents $J^\mu_A$ for shift symmetry can  be written
as 
\ba
&& J^0_\mu = 2 \, U_X \,  \nabla_\mu \varphi^0 + U_Y \, u_\mu + 2
\sum_{m=0}^3 U_{y_m} \, \left( \boldsymbol{B}^m \right)^{ab} \, \nabla_\mu\varphi^b \,
C^{a0} \, ;\\[.2cm]
&& J_\mu^a = \left( b \, U_b  - Y \, U_Y  \right) \left( \boldsymbol{B}^{-1} \right)^{ac} 
 \nabla_\mu \varphi^c + U_Y \, \ell_\mu^a +2 \sum_{m=1}^3
m \, U_{\tau_m} \, \left( \boldsymbol{B}^{m-1} \right)^{ac} \nabla_\mu
\varphi^c \nb\\
&&+2 \, \sum_{m=0}^3 U_{y_m} \sum_{n=1}^3  
\left(\boldsymbol{B}^{m-n} \boldsymbol{Z} \boldsymbol{B}^{n-1}
\right)^{ac} \nabla_\mu \varphi^c + 2 \sum_{m=0}^3 U_{y_m}C^{0b} 
\left( \boldsymbol{B}^{n} \right)^{ba}\nabla_\mu \varphi^0     \;   
\ea
where
\be
\ell^\mu_a = \frac{\epsilon^{\mu \nu \alpha \beta}}{2 \, b \, \sqrt{g}}  \nabla_\nu \varphi^0
  \nabla_\alpha \varphi^a  \nabla_\beta \varphi^b \, \, \epsilon_{abc}
  \, .
\ee
It is also of particular interest the projection of the currents
orthogonal to $u^\mu$, namely $\sigma^\mu_A = h^\mu_\nu J^\nu_A$; we
have
\bea
&& \sigma_\mu^0 = -2 \, U_X \,  \xi_\mu+ 2
\sum_{m=0}^3 U_{y_m} \, \left( \boldsymbol{B}^m \right)^{ab} \, \nabla_\mu \varphi^b \,
C^{a0} \, ; \label{sigma0}\\[.2cm]
&& \sigma_\mu^a= \left( b \, U_b  - Y \, U_Y  \right) \left( \boldsymbol{B}^{-1} \right)^{ac} 
 \, \nabla_\mu \varphi^c + U_Y \, h_\mu^\nu \ell_\nu^a +2 \sum_{m=1}^3
m \, U_{\tau_m} \, \left( \boldsymbol{B}^{m-1} \right)^{ac} \nabla_\mu
\varphi^c \nb \\
&&-2 \, \xi_\mu  \, \sum_{m=0}^3 U_{y_m} \,
\left( \boldsymbol{B}^m 
\right)^{ac} \, C^{c0} + 2 \sum_{m=0}^3 U_{y_m} \,\sum_{n=1}^3  
\left(\boldsymbol{B}^{m-n} \boldsymbol{Z} \boldsymbol{B}^{n-1}
\right)^{ac} \nabla_\mu \varphi^c \; .\label{sigmaa} 
\ea
\section{Masses}
\label{massesapp}
The explicit values of the masses $M_i$, in terms of $U$ and its
derivatives, are given by
\be
\label{MXY}
\begin{split}
& M_0= \frac{\phi'^2}{2 \, \plm^2}
 \left[a^2 \left(U_{Y^2}-2\;
   U_X\right)-4 \;a\; \phi '\; U_{XY}+4 \;U_{X^2} \, \phi'^2\right] \,
,  \qquad M_1= \frac{2\, \phi'^2}{\plm^2} \left(\sum_{n=0}^3 \;  a^{-2 \;n}\;
   U_{y_n}+a^2\;  U_{X}\right),\\
 & M_2= - \frac{2}{\plm^2}\; \sum_{n=1}^3\;  n^2\,a^{2
   (2-n)} \;U_{\tau _n},\\
&M_3=\frac{1}{\plm^2} \left( 
     2  \,\sum_{m,n=1}^3 m\, n\,
   a^{-2 m-2 n+4} \;U_{\tau _m \tau
   _n}+2  \; \sum_n\;n\;  a^{1-2 n}\;
   U_{b \tau _n}- \;
    \sum_{n,m=1}^3 
   \; a^{4-2 n}\; U_{\tau _n}+
   \frac{1}{2\,a^2}\;U_{b^2}  \right) \, ,\\
&  M_4= \frac{1}{\plm^2} \left[\frac{\phi'}{2} \;\left(2\;  \sum_{m,n=1}^3 
   a^{3-2 n}\; U_{Y \tau
   _n}-a^3\; U_Y+   U_{bY}\right)+
  \phi'^2 \;\left(-2\;  \sum_{n=1}^3\,  a^{2-2 n}
   \;U_{X \tau _n}+ a^2\; U_{X}- \frac{1}{a}\,
   U_{b X}\right) \right]\, ;
\end{split}
\ee
where (\ref{bkg}) have been used.
\section{Operators Expansion around FRW}
First order expansion of the fundamental operators
\ba\label{byx}
&& b=\frac{1}{a^3}\;\left(1-3\;\Phi+\partial^2\pi_L \right),\quad
Y=\frac{\phi'}{a}\;\left(1+\Psi+\frac{\pi_0'}{\phi'} \right),\\ \nonumber
&& X =-\left(\frac{\phi'}{a}\right)^2\;\left(1+2\;\Psi+2\;\frac{\pi_0'}{\phi'} \right),\quad
\tau_n=\frac{1}{a^{2\,n}}\;\left[3+2\,n\,(-3\;\Phi+\partial^2\pi_L )\right]\\ \nonumber
&& u_\m=\left(-a+a\;\Psi,\;-a\;\partial_i\pi_L'\right),\quad
\vel_\mu=\left(-a+a\;\Psi,\;-\frac{a}{\phi'}\;\partial_i\pi_0\right)
\ea while the $y_n$ are all of order two.

\section{Equations of Motion}
\label{sec:eqm}
The functions ${\cal F}_i$ entering in the equations for $\Phi$ and
$\delta \sigma$ are given by
\ba
{\cal F}_1 &=& \frac{2 \, M_2 \, \mathcal{H} \left[3 (w-
    c_s^2)+2\right]-2 \, 
   M_2'}{{\bf D}} \, ;
\\
{\cal F}_2 &=&\frac{\phi' \left[3 a^2  (w+1)
 \, \mathcal{H} \, 
   M_2'-2 \, M_2^2 \right]-a^2 M_2 (w+1) \phi' \left[3\, \mathcal{H}^2\left(3
   (w- c_s^2) +2\right)+  \, k^2  \right]  }{18 \, a^4  \,  \plm^2\mathcal{H}^2 \,  (w+1) \;
{\bf D}}
\\
{\cal F}_3 &=& \frac{M_2^2 \, f_2 + M_2 \, f_1 + M_2' \, f_0  }{36 \, a^2
  \plm^4 \,  \mathcal{H}^2 \,  (w+1) \left[a^2 \, k^2 \, (w+1)+2 \, M_2\right]} \, ;\\
{ \cal F}_4 &=&-\frac{ 2\, k^4\, \, M_2 \, \phi ' \left[
    3\, \mathcal{H}^2(3 w+5)+2 \, k^2
  \right]}{27 \mathcal{H}^4 \, (w+1) \;{\bf D}} \, ;
\\
{\cal F}_5 &=&\frac{ k^4 \,  M_2 \, \phi'{}^2}{27 \, a^2\, \plm^2\mathcal{H}^4\, (w+1) \;{\bf D}}\, .
  \ea
  where
  \be
  \begin{split}
& f_2 =72 \, \plm^4 \,\mathcal{H}^2 \,  (w+1)+16 \, \plm^4 \, k^2 \, ;\\[.2cm]
& f_1 = 4a^2 \, (w+1) \plm^4\left \{ \, 3\, \mathcal{H}^2 \; \left[3\, \mathcal{H}^2(3 \,w+5) \left(
   3 \, w-3 \, c_s^2+2\right)+ \, k^2 \,  \left(9
  \,  w+7 -6 \, c_s^2\right)\right]+2 \, k^4 \, \right \} \, ;
\\
&f_0 =- 12 \, a^2 \, \plm^4 \,  (w+1) \, \mathcal{H} \left[3\, \mathcal{H}^2\;
  (3 w+5)+2 \, k^2 \,  \right]\, ;
\end{split}
\ee
with ${\bf D}\equiv\left[a^2 \, k^2 \,
   (w+1)+2 \, 
   M_2\right]$.  
The equation for $\Phi$  has a  smooth  limit when some of the $M_i$
are sent  to zero. 
On the contrary, the equation  for $\delta \sigma$ is singular when
$M_1$ or $M_0$ are sent to zero, see Section \ref{DoFsec} for such a cases. 

\section{Masses for Special Supersolids}
Consider a Lagrangian that depends the operators $w_n$; see \eq{Wn}.
The parameters $\{M_i\}$ can be derived by the following identification
in the formulas \eq{MXY}
\ba
U_{\tau_n}\to U_{w_n},\qquad U_{y_{n}}\to \frac{(n+1)\;a^2}{(\phi')^2}\;U_{w_{n+1}}.
\ea
For special supersolids described by $U(b,\,Y,\,X,\,w_n )$, the
expression \eq{m1eff} for $M_1^{eff}$ takes the form
\ba\label{M1wn}
 M_1^{eff}= a\;\frac{a^2\;\phi'\;U_Y-\;U_b}{\plm^2} \, \, .
\ea
In particular, for the Lagrangian of  the form $U(b\;Y,\;X,\;w_n)$ we have
that $M_1^{eff}=0$ and then only 3 degrees of freedom propagate  \cite{Comelli:2014xga,Comelli:2013txa,Comelli:2013paa,Comelli:2012vz,Bebronne:2007qh}, see Table \ref{tab:prop}.
In presence of the ${\cal O}_{\a\b n}$ operators, the $\{M_i\}$ can be computed by the following substitutions
\ba
 && U_X\to(-1)^{\alpha+1 }\; \alpha\; 
    \bar Y^{-2 (\beta +1)}\;
   y_n^{\beta }\;U_{\mathcal{O}} \, ;\\
   && U_Y\to2 \;(-1)^{\alpha+1 }\;
   (\alpha +\beta ) \;
   \bar Y^{-2 \beta -1} \;y_n^{\beta
   }\;U_{\mathcal{O}} \, ;\\
&&  U_{y_n}\to(-1)^{\alpha } \;\beta\; 
    \bar Y^{-2 \beta }\;
   y_n^{\beta -1}\;U_{\mathcal{O}} \, ;
 \ea
 \ba
 && U_{XX}\to(-1)^{\alpha } \;\alpha\; \bar Y^{-4
   \beta -4}\; y_n^{\beta }\;
   \left[(-1)^{\alpha }\; \alpha\; 
   U_{\mathcal{O}^2} \;y_n^{\beta
   }+(\alpha -1)\; U_{\mathcal{O}} \;\bar Y^{2
   \beta }\right] \, ;\\
&&  U_{XY}\to2 \;(-1)^{\alpha }\;
   \alpha \; (\alpha +\beta )\; \bar Y^{-4
   \beta -3} \;y_n^{\beta }\;
   \left[(-1)^{\alpha }\;
   U_{\mathcal{O}^2}\; y_n^{\beta
   }+U_{\mathcal{O}} \;\bar Y^{2 \beta
   }\right] \, ;
 \\ 
&&U_{YY} \to 2 \;(-1)^{\alpha } \;(\alpha
   +\beta ) \;\bar Y^{-4 \beta -2}\;
   y_n^{\beta }\; \left[2 \;(-1)^{\alpha
   } \;(\alpha +\beta )
   \; U_{\mathcal{O}^2} \; y_n^{\beta }+(2\;
   \alpha +2 \;\beta +1)\;
   U_{\mathcal{O}}\; \bar Y^{2 \beta
   }\right]  ;
  \ea
 where we have  used only that $\bar X=-\bar Y^2$ but not he fact that
 $\bar y_n=0$.  
In the case of the Lagrangian  $U({\cal O}_{\a\b n},\,\tau_n)$ we have 
  \ba\label{myn}
  M_0=\frac{2}{\plm^2} \;(-1)^{\alpha } \;a^2\; \beta \; \bar Y^{-4
   \beta } \; y_n^{\beta } \;\left[2\;
   (-1)^{\alpha } \;\beta\; 
   U_{\mathcal{O}^2}\; y_n^{\beta }+(2\;
   \beta +1)\; U_{\mathcal{O}}\; \bar Y^{2
   \beta }\right]=0 \,  ;
  \ea
irrespective of the value of $\beta$.  Thus, both  when $\beta=0$ and only
the operators ${\cal O}_\a={\cal O}_{\a 0 n}$ are present  (see section \ref{sect:superfluids} and section
\ref{sect:supersolids}) and also for $\b> 0$, due to the fact that $\bar y_n=0$. 

 \end{appendix}

\bibliographystyle{hunsrt}  
  
\bibliography{fluidbiblio}

\end{document}